\documentclass[12pt]{article}
\usepackage{amsfonts}
\usepackage{amssymb}
\usepackage{amsmath}
\usepackage{latexsym}
\usepackage{epsfig}
\usepackage{dsfont,eucal,mathrsfs}

\newif{\ifcomentarios}
\comentariosfalse

\newtheorem{theorem}{Theorem}

\newtheorem{definition}[theorem]{Definition}
\newtheorem{proposition}[theorem]{Proposition}

\newtheorem{remark}[theorem]{Remark}

\renewcommand{\mathbf}{\boldsymbol}
\renewcommand{\mathcal}{\mathscr}

\begin{document}

\author{\textbf{William R. P. Conti}\thanks{
Supported by FAPESP under grant $\#98/10745-1$. E-mail: \textit{\
william@if.usp.br} }\ \ and \ \ \textbf{Domingos H. U. Marchetti}\thanks{
Partially supported by CNPq and FAPESP. E-mail: \textit{\ marchett@if.usp.br}
} \\
Instituto de F\'{\i}sica\\
Universidade de S\~{a}o Paulo \\
Caixa Postal 66318\\
05315 S\~{a}o Paulo, SP, Brasil }
\title{Hierarchical Spherical Model as a Viscosity Limit of Corresponding $%
O(N)$ Heisenberg Model}
\date{}
\maketitle

\begin{abstract}
The $O(N)$ Heisenberg and spherical models with interaction given by the
long range hierarchical Laplacean are investigated. Two classical results
are adapted. The Kac--Thompson solution \cite{KT} of the spherical model,
which holds for spacially homogeneous interaction, is firstly extended to
hierarchical model whose interaction fails to be translation invariant.
Then, the convergence proof of $O(N)$ Heisenberg to the spherical model by
Kunz and Zumbach \cite{KZ} is extended to the long range hierarchical
interaction. We also examine whether these results can be carried over as
the size of the hierarchical block $L$ goes to $1$. These extensions are
considered a preliminary study prior the investigation of the model by
renormalization group given in \cite{MCG} where central limit theorems for
the spherical ($N=\infty $) model on the local potential approximation ($%
L\downarrow 1$) are then established from an explicit solution of the
associate nonlinear first order partial differential equation.
\end{abstract}

\section{Introduction}

\setcounter{equation}{0} \setcounter{theorem}{0}

\noindent \textit{Motivations. }In the present work we initiate a geometric
study of partial differential equations related to the renormalization group
transformation (RGT) of a $d$--dimensional $N$--component hierarchical spin
system in the limit as the block size $L^{d}$ goes to $1$.

For one--component hierarchical spin system, the evolution equation
corresponding to $L\downarrow 1$ limit, known as the local potential
approximation, began to be investigated by Felder \cite{F} who has
constructed, in that seminal work, a family of global stationary solutions
-- nontrivial fixed point analogs of the corresponding $d$--dimensional
models with $d>2$ and $L^{d}$ an integer $>1$. Except by a preliminary study
and numerical simulation of the corresponding $N$--component spin equation
by Zumbach \cite{Z} and a refinement of Felder's approach by Lima \cite{L},
no much attention has been paid to the evolution equation model.

Very recently, the renormalization group trajectories of $d=4$ dimensional
hierarchical $O(N)$ spin model with block size $L^{d}=2$ has been
investigated by Watanabe \cite{W} (for previous investigations, see \cite%
{GK,K} and references therein). Starting from the uniform \textquotedblleft
a priori\textquotedblright\ measure supported in the $N$--dimensional sphere
of radius $\sqrt{N}$, the critical trajectory has shown to converge to the
Gaussian fixed point for sufficiently large $N$. To control such trajectory,
which starts far away from the fixed point, the exactly solved $O(\infty )$
trajectory has been used together with two key ingredients: reflection
positivity and the Lee--Yang property of\ single--site\ \textquotedblleft a
priori\textquotedblright\ measures. The former ingredient gives uniform
convergence of $O(N)$ trajectories to $O(\infty )$ trajectories. The latter
property has been previously employed by Kozitsky \cite{K} to establish
central limit theorems for the hierarchical $O(N)$ spin model with block
size $L^{d}\geq 2$ and $d>4$ (in his notation $L^{d}=\delta $ and $%
2/d=\lambda \in \left( 0,1/2\right) $) at the critical inverse temperature $%
\beta _{c}$ and below. Watanabe's analysis, based in his joint work with
Hara and Hatttori \cite{HHW} on the critical trajectory for the hierarchical
Ising model ($N=1$), in contradistinction, deals with the borderline $d=4$
case and does not require closeness to the fixed point.

Although the analysis of the RGT with $L^{d}\geq 2$ fixed is expected to be
simplified considerably in the $L\downarrow 1$ limit, none of the results on
the critical trajectories can be carried to the limit as the above mentioned
ingredients do not hold if $L^{d}$ is not an integer. To establish weak
convergence of the hierarchical $O(N)$ Heisenberg equilibrium measure to the
corresponding spherical equilibrium measure as $N\rightarrow \infty $ in the
local potential approximation ($L\downarrow 1$) an entirely new method of
analysis has to be developed from scratch.

The present investigation establishes some classical results on the $O(N)$
Heisenberg and spherical model with short range discrete Laplacean
interaction replaced by\ the long range hierarchical Laplacean and examines
whether they can be carried over as $L$ goes to $1$. Kac--Thompson solution
of the spherical model \cite{KT}, which holds for spatially homogeneous
interaction, is presented in Section \ref{SM} and extended in Section \ref%
{HSM} to hierarchical model whose interaction fails to be translation
invariant. Kac--Thompson's asymptotic analysis is then applied to the moment
generating function of the block spin random variable and normal
fluctuations are established for the spherical model for $\beta <\beta _{c}$%
. To our knowledge, such application and consequences are new. In Section %
\ref{CSM} the convergence proof of $O(N)$ Heisenberg to the spherical model
by Kunz and Zumbach \cite{KZ} is extended to the long range hierarchical
interaction. The prove holds for the free energy and the moments generating
function. These extensions are considered a preliminary study prior the
investigation of the model by renormalization group. In a subsequent work 
\cite{MCG} we establish central limit theorems for the spherical model on
the local potential approximation from an explicit solution of the associate
nonlinear first order partial differential equation given by (\ref{nleq})
with $N=\infty $.

\smallskip

\noindent \textit{Viscosity Limit. }The hierarchical Heisenberg model on a
box $\Lambda \subset \mathbb{Z}^{d}$ of size $n=L^{dK}$ is defined by an $%
O(N)$ invariant equilibrium measure%
\begin{equation}
d\nu _{n}^{(N)}(\mathbf{x})=\frac{1}{Z_{n}}\exp \left\{ -\frac{1}{2}\left( 
\mathbf{x},A\mathbf{x}\right) \right\} \prod_{j=1}^{n}d\sigma
_{0}^{(N)}\left( x_{j}\right)  \label{equ}
\end{equation}%
where $\mathbf{x}=\left( x_{1},\ldots ,x_{n}\right) $ denotes an element of
the configuration space $\Omega _{n}=\mathbb{R}^{N}\times \cdots \times 
\mathbb{R}^{N}$, $A=J\otimes I$ the tensor product of the $n\times n$
coupling hierarchical matrix $J$ (see (\ref{J})) with the $N\times N$
identity matrix $I$ and $\sigma _{0}\left( x\right) $ the a priori uniform
measure on the $N$--dimensional sphere $\left\vert x\right\vert ^{2}=\beta N$
of radius $\sqrt{\beta N}$.

The invariance of $J$ under the block spin transformation (\ref{B}) allows
to describe the laws of equilibrium studying the dynamics of a recursion
relation 
\begin{equation*}
\sigma _{k}(x)=\mathcal{R}\sigma _{k-1}(x)
\end{equation*}%
in the space of single--site \textquotedblleft a priori\textquotedblright\
measures, which in terms of their characteristic functions 
\begin{equation}
\phi _{k}^{(N)}(z)=\int d\sigma _{k}^{(N)}(x)~\exp \left( iz\cdot x\right) ~,
\label{cf}
\end{equation}%
reads%
\begin{equation}
\phi _{k}^{(N)}(z)=\frac{1}{N_{k}}\exp \left( \frac{-1}{2}\Delta \right)
\left( \phi _{k-1}^{(N)}(L^{-\gamma /2}z)\right) ^{L^{d}}  \label{rr}
\end{equation}%
with $\gamma =d+2$ and initial condition%
\begin{equation}
\phi _{0}^{(N)}(z)=\frac{J_{N/2-1}\left( \sqrt{\beta N}\left\vert
z\right\vert \right) }{\left( \sqrt{\beta N}\left\vert z\right\vert
/2\right) ^{N/2-1}}\Gamma \left( N/2\right) =\varphi _{0}^{(N)}(\left\vert
z\right\vert )  \label{ic}
\end{equation}%
Here, $\exp \left( t\Delta \right) $ is the semi--group generated by the $N$%
--dimensional Laplacean $\Delta =\partial ^{2}/\partial z_{1}^{2}+\cdots
+\partial ^{2}/\partial z_{N}^{2}$, $N_{k}$ is chosen so that $\phi
_{k}(0)=1 $ holds for all $k=1,\ldots ,K$ and $J_{\alpha }(x)$ is the Bessel
function of order $\alpha $. Note that $\phi _{k}^{(N)}(z)=\varphi
_{k}^{(N)}(\left\vert z\right\vert )$ depends only on the norm $\left\vert
z\right\vert ^{2}=z\cdot z$ and $\beta $ is the inverse temperature.

We shall now explain our title. Since $\psi (t,z)=\exp \left( t\Delta
\right) \psi _{0}(z)=\exp (-U(t,z))$ satisfies the heat equation with
initial condition $\psi (0,\cdot )=\psi _{0}$, $U$ satisfies $U_{t}-\Delta
U+\left\vert \nabla U\right\vert ^{2}=0$ with $U(0,\cdot )=U_{0}=-\ln \psi
_{0}$ and the sequence $u_{k}^{(N)}(x)$, $k=0,1,\ldots $, defined (with $%
\left\vert z\right\vert ^{2}=-Nx$) by 
\begin{equation*}
\varphi _{k}^{(N)}(\sqrt{-Nx})=\exp \left( -Nu_{k}^{(N)}(x)\right) 
\end{equation*}%
can be obtained by solving a nonlinear heat equation\footnote{%
Subindex $t$ and $x$ refer to partial derivatives with respect to
independent variables.}%
\begin{equation}
u_{t}-\frac{2}{N}xu_{xx}-u_{x}+2xu_{x}^{2}=0  \label{heat}
\end{equation}%
up to time $t=1/2$ starting from the initial conditions $%
u(0,x)=L^{d}u_{k-1}^{(N)}(L^{-\gamma }x)$:%
\begin{equation}
u_{k}^{(N)}(x)=u(1/2,x)-u(1/2,0)~  \label{init}
\end{equation}%
(the solution at $x=0$ is subtracted to satisfy $\varphi
_{k}^{(N)}(0)=e^{-Nu_{k}^{(N)}(0)}=1$). This is the starting point of
Watanabe's investigation \cite{W} (see also \cite{HHW}) on Taylor
coefficients $\nu _{2l,k}^{(N)}$, $l\geq 1$, of the sequence of functions $%
v_{k}^{(N)}(\zeta )=-\ln \varphi _{k}^{(N)}(\sqrt{N}\zeta )/N$ around $\zeta
=0$. As already recognized by Watanabe, the parabolic equation (\ref{heat})
becomes a first order hyperbolic equation when $N\rightarrow \infty $ and
the expansion in powers of $1/N$ of a trajectory $\left( u_{k}^{(N)}\right)
_{k\geq 0}$ for the $N$--vectorial hierarchical model is a singular
perturbation about the corresponding trajectory for the spherical
hierarchical model.

The local potential approximation replaces the exponent $1/2$ in (\ref{rr})
by $(L-1)/2$ and the interval of time evolved by (\ref{heat}) tends to $0$
when $L\downarrow 1$. As a consequence, defining $%
u^{(N)}(t,x)=u_{k}^{(N)}(x) $ for $t=k\ln L$ and taking the limit $%
L\downarrow 1$, $k\rightarrow \infty $ with $t$ fixed, the recursive initial
value problem (\ref{heat}) and (\ref{init}) for $\left( u_{k}^{(N)}\right)
_{k\geq 0}$ turn into a genuine initial value problem given by%
\begin{equation}
u_{t}^{(N)}-\frac{2}{N}xu_{xx}^{(N)}-u_{x}^{(N)}+2x\left( u_{x}^{(N)}\right)
^{2}+\gamma xu_{x}^{(N)}-du^{(N)}+u_{x}^{(N)}(t,0)=0  \label{nleq}
\end{equation}%
with $u^{(N)}(0,x)=-\ln \varphi _{0}^{(N)}(\sqrt{-Nx})/N\equiv u_{0}^{(N)}$.
Comparing to (\ref{heat}), (\ref{nleq}) includes three extra terms, the last
one ensures $u^{(N)}(t,0)=0$ for all $t\geq 0$, corresponding to the
operations of dilation, multiplication and normalization performed between
two consecutive evolutions of (\ref{heat}). Note that the stationary
solution $u^{\ast }(x)=-x$ of (\ref{nleq}) corresponds to the Gaussian fixed
point of (\ref{rr}). In the second work of our series \cite{MCG} we give a
geometric description of the trajectory $\left\{ u^{(\infty )}(t,x),t\geq
0\right\} $, in the viscosity limit $N=\infty $,\footnote{$1/N$ plays the
role of viscosity since it is in front of the Laplacean as in the
hydrodynamic equation of incompressible fluid.} at and above the critical
inverse temperature. Our third investigation will address the solution $%
\left\{ u^{(N)}(t,x),t\geq 0\right\} $ of (\ref{nleq}) as a (singular)
perturbation about the critical trajectory $\left\{ u^{(\infty )}(t,x),t\geq
0\right\} $.

\smallskip

\noindent \textit{Statement of Results. }Equilibrium laws of the model are
described by the distribution of block spin random variable $X_{n}^{\gamma
}=n^{-\gamma /(2d)}\sum_{j=1}^{n}x_{j}$ in the limit as $n\rightarrow \infty 
$ with a properly chosen $\gamma $. The characteristic function with respect
to the equilibrium measure (\ref{equ}) of the block variable with $\gamma
=d+2$ reads 
\begin{eqnarray*}
\Phi _{n}^{(N)}\left( z\right)  &=&\int \exp \left(
iL^{-K(d+2)/2}\sum_{j=1}^{n}x_{j}\cdot z\right) d\nu _{n}^{(N)}(\mathbf{x})
\\
&=&\int \exp \left( ix\cdot z\right) ~d\sigma _{K}^{(N)}(x)=\varphi
_{K}^{(N)}(\left\vert z\right\vert )~.
\end{eqnarray*}%
The equilibrium distribution $\sigma _{K}^{(N)}(x)=\nu
_{n}^{(N)}(X_{n}^{\gamma }\leq x)$ converges weakly in the thermodynamic
limit $n=L^{dK}\rightarrow \infty $ to $\sigma ^{(N)}(x)=\nu
^{(N)}(X^{\gamma }\leq x)$ if $\varphi _{K}^{(N)}(\left\vert z\right\vert )$
is continuous at origin and converges pointwise to a continuous (at origin)
function $\varphi ^{(N)}(\left\vert z\right\vert )$. Hence, the equilibrium
distribution $\nu ^{(N)}(X^{\gamma }\leq x)$ converges weakly to the
equilibrium measure of the spherical model if $\lim_{N\rightarrow \infty
}\left( \varphi ^{(N)}(\sqrt{N}\left\vert z\right\vert )\right)
^{1/N}=\varphi ^{(\infty )}(\left\vert z\right\vert )$ exist for every point 
$z$ and coincides with the corresponding characteristic function of the
latter model provided $\varphi ^{(\infty )}(\left\vert z\right\vert )$ is
continuous at $z=0$. These statements, which is independent of which order
the limits $n\rightarrow \infty $ and $N\rightarrow \infty $ are taken, are
proven in Section \ref{CSM} for $\gamma =d$, $L^{d}\geq 2$ integer and $%
\beta $ different from the critical inverse temperature $\beta _{c}=\beta
_{c}(d,L)$ of the spherical model.

The following result on the moment generating function holds for admissible
coupling matrices, including hierarchical matrix. 

\begin{theorem}
\label{convergence1}The finite volume moment generating function of the
Heisenberg model%
\begin{equation*}
\Theta _{n}^{(N)}(\beta ,z)=\int \exp \left( \frac{z}{\sqrt{nN}}%
\sum_{i=1}^{n}\sum_{j=1}^{N}x_{i,j}\right) d\nu _{n}^{(N)}(\mathbf{x})
\end{equation*}%
with admissible reflection positive sequence of coupling matrices $A$,
converges%
\begin{equation}
\lim_{n,N\rightarrow \infty }\Theta _{n}^{(N)}(\beta ,z)=\Theta (\beta ,z)
\label{Theta}
\end{equation}%
to the spherical model moment generating function $\Theta (\beta ,z)$ (see (%
\ref{FF})) as $n$, $N$ goes to infinity in any order, for $\beta <\beta
_{c}(A)$ given by (\ref{critical}) and uniformly in compact intervals of $%
z\in \mathbb{R}$.
\end{theorem}

\section{Spherical Model\label{SM}}

\setcounter{equation}{0} \setcounter{theorem}{0}

We review the solution and some basic properties of Berlin--Kac model with a
positive definite coupling matrix $A$ satisfying a condition stated in (\ref%
{lim}). Formulas written in this section are independent on whether
translational invariance holds and are, in addition, suitable to the
hierarchical coupling matrix investigated in the next section. We shall make
most of those expressions explicit by choosing $A$ the usual discrete
Laplacean, denoted here by $-\Delta $. The same symbol will be used for the
hierarchical Laplacean in Section \ref{HSM}.

\subsection{The Free Energy \label{SFE}}

Given $\beta \geq 0$ and a positive coupling matrix $J=[J_{ij}]_{i,j=1}^{n}$%
, the spherical model of Berlin and Kac \cite{BK} associated with $\beta $
and $J$ is defined by the partition function%
\begin{equation}
Q_{n}(\beta ,J)=\frac{1}{S_{n}}\int d\sigma _{n}(\mathbf{x};\sqrt{n})~\exp
\left\{ \frac{-\beta }{2}\left( \mathbf{x},J\mathbf{x}\right) \right\}
\label{Qn}
\end{equation}%
where $\left( \mathbf{x},\mathbf{y}\right) =\displaystyle%
\sum_{i=1}^{n}x_{i}~y_{i}$ denotes the inner product in $\mathbb{R}^{n}$, 
\begin{equation}
d\sigma _{n}(\mathbf{x};r)=\delta \left( \left\Vert \mathbf{x}\right\Vert
-r\right) \prod_{i=1}^{n}dx_{i}  \label{r}
\end{equation}%
the uniform measure on the sphere $\Sigma _{n}(r)=\left\{ \mathbf{x}\in 
\mathbb{R}^{n}:\left\Vert \mathbf{x}\right\Vert ^{2}=\left( \mathbf{x},%
\mathbf{x}\right) =r^{2}\right\} $ of radius $r$ and 
\begin{equation}
S_{n}=\int d\sigma _{n}(\mathbf{x};\sqrt{n})=\frac{2\pi ^{n/2}n^{(n-1)/2}}{%
\Gamma (n/2)}  \label{Sn}
\end{equation}%
is the surface area of the sphere $\Sigma _{n}\left( \sqrt{n}\right) $.

The most common choice of coupling matrix $J$ is given by the discrete
Laplacean, $-\Delta _{\Lambda }$,\footnote{%
For simplicity, we drop the subindex of Laplacean $\Delta _{\Lambda }$ if no
confusion exists.} on a $d$--dimensional hypercube $\Lambda \subset \mathbb{Z%
}^{d}$ of size $n=L^{d}$ with periodic boundary condition: 
\begin{equation}
-\left( \Delta _{\Lambda }f\right) _{i}=\sum_{j:\left\vert i-j\right\vert
=1}\left( f_{i}-f_{j}\right) =2df_{i}-\sum_{j:\left\vert i-j\right\vert
=1}f_{j}~  \label{laplacean}
\end{equation}%
with the summation over the lattice sites $j$ which are at unit Euclidean
distance from $i\in \Lambda $.

\begin{remark}
The Berlin--Kac model incorporates essential features of the ferromagnetic
Ising model, exhibits a phase transition and has the advantage to be exactly
solvable in any dimension. See \cite{BK} for an extensive discussion on the
thermodynamic properties above and below the critical temperature. The phase
transition on the spherical model is of the same nature of that observed in
the free Bose gas in which condensation of a single mode occurs ( see e.g. 
\cite{P}). Disordered mean spherical model and equivalence of ensembles has
been investigated by Pastur (see \cite{KKPS} and references therein). See
also Perez--Wreszinski--van Hemmen \cite{PWH}\ for disordered spherical
models.
\end{remark}

To solve the spherical model it is convenient to introduce an auxiliary
expression 
\begin{eqnarray}
I_{n} &=&\frac{1}{S_{n}}\int_{\mathbb{R}^{n}}\prod_{i=1}^{n}dx_{i}~\exp
\left\{ \frac{-1}{2}\left( \mathbf{x},\left( J-\mu \right) \mathbf{x}\right)
\right\}  \notag \\
&=&\frac{1}{S_{n}}\int_{0}^{\infty }dr\int d\sigma _{n}(\mathbf{x};r)~\exp
\left\{ \frac{-1}{2}\left( \mathbf{x},\left( J-\mu \right) \mathbf{x}\right)
\right\}  \label{In}
\end{eqnarray}%
which, after changing variables $r=\sqrt{n}s$ and $\mathbf{x}=s\mathbf{y}$,
can be written as%
\begin{equation}
I_{n}=\sqrt{n}\int_{0}^{\infty }\frac{ds}{s}~\exp \left\{ n~h_{n}(s)\right\}
\label{I}
\end{equation}%
where%
\begin{equation}
h_{n}(s)=\frac{\mu }{2}s^{2}+\ln s+f_{n}(s^{2})  \label{h}
\end{equation}%
and

\begin{equation}
f_{n}\left( s^{2}\right) =\frac{1}{n}\ln Q_{n}\left( s^{2},J\right)
\label{fn}
\end{equation}%
is the finite volume free energy of the spherical model.

$I_{n}$ may be think as the grand--canonical partition function with the
Lagrange multiplier $\mu <0$ playing the role of a chemical potential. The
function $I_{n}$ can be integrated%
\begin{equation}
I_{n}=\frac{2^{n/2-1}~\Gamma (n/2)}{n^{(n-1)/2}\sqrt{\det \left( J-\mu
\right) }}  \label{I2}
\end{equation}%
and equations (\ref{I}) and (\ref{I2}) used to evaluate the free energy when 
$n\rightarrow \infty $.

For instance, if $J$ is given by (\ref{laplacean}), Fourier spectral
analysis (see \cite{D}) can be used in (\ref{I2}) together with Stirling's
formula $\Gamma (n/2)\sim \left( n/2e\right) ^{n/2}$ to write 
\begin{equation}
\lim_{n\rightarrow \infty }\frac{1}{n}\ln I_{n}=-\frac{1}{2}-\frac{1}{2}%
\mathbb{E}\ln \left( -\Delta -\mu \right) ~  \label{I1}
\end{equation}%
where 
\begin{eqnarray}
\mathbb{E}\ln \left( -\Delta -\mu \right) &=&\lim\limits_{L\rightarrow
\infty }\dfrac{1}{L^{d}}\sum_{\substack{ m\in \mathbb{Z}^{d}:  \\ %
-L/2<m_{l}\leq L/2}}\ln \left( \omega \left( 2\pi m/L\right) -\mu \right) 
\notag \\
&=&\frac{1}{\left( 2\pi \right) ^{d}}\int_{[-\pi ,\pi ]^{d}}d^{d}k~\ln
\left( \omega (k)-\mu \right) ~  \label{EE}
\end{eqnarray}%
and%
\begin{equation*}
\omega (k)=4\sum_{l=1}^{d}\sin ^{2}\frac{k_{l}}{2}~.
\end{equation*}%
Note that, as $n\rightarrow \infty $, $-\Delta $ is unitarily equivalent to
an operator of multiplication by $\omega (k)$ in the space $L_{2}\left( %
\left[ -\pi ,\pi \right] ^{d},\mathbb{C}\right) $ of square integrable
functions $f:\left[ -\pi ,\pi \right] ^{d}\longrightarrow \mathbb{C}$.

Let us now state our assumption on $J$ and explain the probabilistic
notation $\mathbb{E}\left( \cdot \right) $ in (\ref{I1}).

\begin{definition}
\label{adm}A sequence $A=\left\{ A_{n}\right\} _{n\geq 1}$ of coupling
matrices ($n$ indicates the order of $A_{n}$) is an admissible sequence if
each $A_{n}$ is nonnegative ($\left( A_{n}\right) _{ij}\geq 0$) positive
definite real symmetric matrix and 
\begin{equation}
\mathbb{E}f(A)\equiv \lim\limits_{n\rightarrow \infty }\dfrac{1}{n}\text{Tr}%
f(A_{n})  \label{lim}
\end{equation}%
exists for every continuous bounded function $f$.

We require, in addition, that $\mathbf{1}=\left( 1/\sqrt{n}\right) \left(
1,\ldots ,1\right) $ is an eigenvector of $A_{n}$ with associate eigenvalue $%
0$.
\end{definition}

From now on, only admissible sequences of coupling matrices will be
considered. By definition, 
\begin{equation}
\mathbb{E}f(A)=\lim\limits_{n\rightarrow \infty }\dfrac{1}{n}%
\sum_{i=1}^{n}f\left( \lambda _{i}^{(n)}\right) =\lim\limits_{n\rightarrow
\infty }\int d\rho _{n}(\lambda )~f(\lambda )=\int d\rho (\lambda
)~f(\lambda )  \label{T}
\end{equation}%
is the expectation with respect to the empirical distribution $\rho $ which
is the weak limit of the integrated density of eigenvalues $\lambda
_{1}^{(n)},\ldots ,\lambda _{n}^{(n)}$ (counting multiplicity) of $A_{n}$:%
\begin{equation}
\rho _{n}(\lambda )=\frac{1}{n}\sum_{i=1}^{n}\chi _{\lbrack \lambda
_{i}^{(n)},\infty )}(\lambda )  \label{rhon}
\end{equation}%
with $\chi _{\lbrack a,b)}(\lambda )=1$ if $\lambda \in \lbrack a,b)$ and $%
\chi _{\lbrack a,b)}(\lambda )=0$ otherwise.

The empirical measure $\rho $ associated with $-\Delta $ is absolutely
continuous with respect to the Lebesgue measure $d\rho (\lambda )=\rho
^{\prime }(\lambda )d\lambda $, 
\begin{equation*}
\rho ^{\prime }(\lambda )=\frac{1}{\pi }\lim_{\varepsilon \downarrow 0}\Im
\left( \left( -\Delta -\lambda -i\varepsilon \right) _{00}^{-1}\right)
\end{equation*}%
exists for almost every $\lambda $ and is supported in the interval $\left[
0,4d\right] $. For $d=1$, we have explicitly $\rho ^{\prime }(\lambda
)=1\left/ \left( \pi \sqrt{4\lambda -\lambda ^{2}}\right) \right. $.

On the other hand, Laplace's asymptotic method applied to (\ref{I}) gives%
\begin{equation}
I_{n}=\sqrt{\frac{2\pi }{-h^{\prime \prime }(\bar{s})}}\frac{1}{\bar{s}}%
~\exp \left\{ n~h(\bar{s})\right\} \left( 1+O\left( 1/n\right) \right)
\label{Itilde}
\end{equation}%
with $\bar{s}$ the value at which $h(s)=\lim\limits_{n\rightarrow \infty
}h_{n}(s)$ attains its maximum value. The existence of a unique strictly
positive maximum $\bar{s}$ follows from certain facts that are independent
of $J$ considered. Writing $Q_{n}(s^{2})=Q_{n}(s^{2},J)$, we have

\begin{enumerate}
\item $Q_{n}(0)=1$

\item $Q_{n}^{\prime }(s^{2})=\dfrac{-1}{2S_{n}}\int d\sigma _{n}(\mathbf{x};%
\sqrt{n})~\left( \mathbf{x},J\mathbf{x}\right) \exp \left\{ \dfrac{-s^{2}}{2}%
\left( \mathbf{x},J\mathbf{x}\right) \right\} $

\item 
\begin{eqnarray*}
0 &\leq &Q_{n}^{\prime \prime }(s^{2})Q_{n}(s^{2})-\left( Q_{n}^{\prime
}(s^{2})\right) ^{2}=\frac{1}{8S_{n}^{2}}\int d\sigma _{n}(\mathbf{x};\sqrt{n%
})~d\sigma _{n}(\mathbf{y};\sqrt{n}) \\
&&\times \left( \left( \mathbf{x},J\mathbf{x}\right) -\left( \mathbf{y},J%
\mathbf{y}\right) \right) ^{2}\exp \left\{ \frac{-s^{2}}{2}\left[ \left( 
\mathbf{x},J\mathbf{x}\right) +\left( \mathbf{y},J\mathbf{y}\right) \right]
\right\}
\end{eqnarray*}
\end{enumerate}

If $J$ satisfies (\ref{lim}) then $\left\Vert J\right\Vert \leq c$ is
bounded (when $J=-\Delta $, for instance, we have 
\begin{equation*}
0\leq \frac{1}{2}\left( \mathbf{x},J\mathbf{x}\right) \leq \frac{1}{2}%
\left\Vert J\right\Vert \left\Vert \mathbf{x}\right\Vert ^{2}=2dn~,
\end{equation*}%
in view of (\ref{r}) and $\left\Vert J\right\Vert =\sup_{k\in \left[ -\pi
,\pi \right] ^{d}}\omega (k)=4d$) and together with the mean value theorem,
we conclude $f_{n}(s^{2})=f_{n}^{\prime }(\tilde{s}^{2})~s^{2}$ holds for
some $0<\tilde{s}<s$, with%
\begin{equation*}
-\frac{c}{2}\leq f_{n}^{\prime }(s^{2})=\frac{1}{n}\frac{Q_{n}^{\prime
}(s^{2})}{Q_{n}(s^{2})}\leq 0
\end{equation*}%
and%
\begin{equation*}
f_{n}^{\prime \prime }(s^{2})=\frac{1}{n}\left( \frac{Q_{n}^{\prime \prime
}(s^{2})}{Q_{n}(s^{2})}-\left( \frac{Q_{n}^{\prime }(s^{2})}{Q_{n}(s^{2})}%
\right) ^{2}\right) \geq 0
\end{equation*}%
uniformly in $n$. $\left\{ f_{n}(s^{2})\right\} $ is a sequence of convex
functions, uniformly bounded in every compact of $\mathbb{R}_{+}$ and there
is a subsequence $\left\{ f_{n_{j}}(s^{2})\right\} $ such that $%
\lim\limits_{j\rightarrow \infty }$ $f_{n_{j}}(s^{2})=$ $f(s^{2})$ exists
with $f(s^{2})$ convex and differentiable at almost every $s^{2}\in \mathbb{R%
}_{+}$ (See \cite{KT} and Section 3.1 of \cite{KKPS}). As a consequence, the
maximum of $h$ is attained at%
\begin{equation}
\bar{s}^{2}=\bar{s}^{2}(\mu )=\frac{-1}{\mu +2f^{\prime }(\bar{s}^{2})}
\label{nu2}
\end{equation}%
for every $\mu \leq 0$.

Equating (\ref{I1}) and (\ref{Itilde}) together with (\ref{h}) yields 
\begin{equation*}
\frac{\mu }{2}\bar{s}^{2}+\ln \bar{s}+f(\bar{s}^{2})=-\frac{1}{2}-\frac{1}{2}%
\mathbb{E}\ln \left( J-\mu \right)
\end{equation*}%
which, by the inverse function theorem, can be differentiated with respect
to $\mu $:%
\begin{equation}
\bar{s}^{2}(\mu )=\mathbb{E}\left( J-\mu \right) ^{-1}  \label{nu}
\end{equation}%
in view of $h^{\prime }(\bar{s})=0$. When $J=-\Delta $, the equation reads 
\begin{equation}
\bar{s}^{2}(\mu )=\frac{1}{\left( 2\pi \right) ^{d}}\int_{[-\pi ,\pi
]^{d}}d^{d}k~\frac{1}{\omega (k)-\mu }~.~~  \label{nu1}
\end{equation}

Solving (\ref{nu}) for $\mu =\mu (\bar{s}^{2})$, with $\bar{s}^{2}$ replaced
by an arbitrary positive number $\beta $, substituting back to the previous
equation gives the free energy of the spherical model%
\begin{equation}
f(\beta )=-\mu (\beta )\frac{\beta }{2}-\ln \left( \sqrt{e\beta }\right) -%
\frac{1}{2}\mathbb{E}\ln \left( J-\mu (\beta )\right)  \label{f}
\end{equation}%
(for $J=-\Delta $ the last term is given by (\ref{EE})).

\begin{remark}
\label{condensate}The sum--rule (\ref{nu}) can be obtained directly from the
grand--canonical partition function (\ref{In}) (see e.g. \cite{P})%
\begin{equation*}
s^{2}=\lim_{n\rightarrow \infty }\frac{2}{n}\frac{\partial \ln I_{n}}{%
\partial \mu }
\end{equation*}%
and this expresses the equivalence between different ensembles of this
model. As a function of $\mu \in (-\infty ,0]$, $H_{n}(\mu )=\left(
2/n\right) \partial \ln I_{n}/\partial \mu $ is convex, monotone increasing
with $H_{n}(-\infty )=0$ and $\lim_{\mu \uparrow 0}H_{n}(\mu )=\infty $. If $%
\mathcal{P}_{0}$ projects on the invariant subspace of $-\Delta $ associated
with $\lambda =0$, it follows \qquad 
\begin{eqnarray}
s_{0}^{2} &\equiv &\mathbb{E}\mathcal{P}_{0}\left( -\Delta -\mu I\right)
^{-1}  \notag \\
&=&s^{2}-\frac{1}{\left( 2\pi \right) ^{d}}\int_{[-\pi ,\pi ]^{d}\backslash
\left\{ 0\right\} }d^{d}k~\frac{1}{\omega (k)-\mu }  \notag \\
&\geq &s^{2}-\frac{1}{\left( 2\pi \right) ^{d}}\int_{[-\pi ,\pi ]^{d}}d^{d}k~%
\frac{1}{\omega (k)}=s^{2}-\mathbb{E}\left( -\Delta \right) ^{-1}  \label{0}
\end{eqnarray}%
The $0$ eigenvalue is said to condensate if $s_{0}^{2}>0$. According to the
above inequality, $s_{0}^{2}>0$ provided $\beta =s^{2}>\mathbb{E}\left(
-\Delta \right) ^{-1}$. Since (\ref{nu}) has a unique solution $\mu =\mu (%
\bar{s})$ with $s_{0}^{2}=0$ for $s^{2}\leq $ $\mathbb{E}\left( -\Delta
\right) ^{-1}$, the spherical model exhibit a phase transition of
Bose--Einstein type whenever $\mathbb{E}\left( -\Delta \right) ^{-1}$ is
finite, i. e. when $d>2$. Note that, for $k$ near $0$, $\omega (k)\sim
\left\vert k\right\vert ^{2}$ and%
\begin{equation*}
\int_{\varepsilon }^{1}\frac{1}{k^{2}}k^{d-1}dk=\left\{ 
\begin{array}{lll}
\left( 1-\varepsilon ^{d-2}\right) /(d-2) & \mathrm{if} & d\neq 2 \\ 
\ln 1/\varepsilon & \mathrm{if} & d=2%
\end{array}%
\right. ~
\end{equation*}%
has no limit for $d\leq 2$. For any sequence $A$ of admissible coupling
matrices, the critical inverse temperature of the spherical model is defined
by%
\begin{equation}
\beta _{c}(A)=\mathbb{E}A^{-1}  \label{critical}
\end{equation}
\end{remark}

\medskip

\subsection{Moments Generating Function\label{SMG}}

We state the main result of this section.

\begin{theorem}
\label{CLT}The block spin random variable 
\begin{equation}
X_{n}=\frac{1}{\sqrt{n}}\sum_{i=1}^{n}x_{i}~.  \label{Xn}
\end{equation}%
of a spherical model with admissible sequence $J=\left( J_{n}\right) _{\geq
1}$ of coupling matrices, converge in distribution to a Gaussian variable of
zero mean and variance $-1/\mu $ where $\mu =\mu (\beta )$ is the solution
of (\ref{nu}) with $\bar{s}^{2}=\beta $.
\end{theorem}

\begin{remark}
Since $\mu $ approaches $0$ from below as $\beta \uparrow \beta _{c}$, the
variance $-1/\mu $ diverges and Theorem \ref{CLT} does not hold anymore.
Theorem \ref{CLT} holds for cases (e.g. for no translational invariant $J$%
's) in which the general result of Newman \cite{N} on Gibbs measure
satisfying FKG property with finite susceptibility $\chi $ cannot be applied
.
\end{remark}

Let%
\begin{equation}
Q_{n}(\beta ,z,\mathbf{h},J)=\frac{1}{S_{n}}\int d\sigma _{n}(\mathbf{x};%
\sqrt{n})~\exp \left\{ \frac{-\beta }{2}\left( \mathbf{x},J\mathbf{x}\right)
+z\left( \mathbf{h},\mathbf{x}\right) \right\} ~  \label{Qnzeta}
\end{equation}%
be the partition function of the spherical model including a magnetic field $%
\mathbf{h}=\left( h_{1},\ldots ,h_{n}\right) $. Since $Q_{n}(\beta ,z,%
\mathbf{h},J)=Q_{n}(\beta ,-z,\mathbf{h},J)$, (\ref{Qnzeta}) is an even
function of $z$ and $Q_{n}(\beta ,0,\mathbf{h},J)=Q_{n}(\beta ,J)$ is given
by (\ref{Qn}). One easily verifies that the ratio of partition functions 
\begin{equation}
\Theta _{n}(\beta ,z)=\frac{Q_{n}(\beta ,z,\mathbf{1},J)}{Q_{n}(\beta ,J)}
\label{thetan}
\end{equation}%
with $\mathbf{h}$ in (\ref{Qnzeta}) given by the $n$--component unit vector $%
\mathbf{1}=\left( 1/\sqrt{n}\right) \left( 1,\ldots ,1\right) $, generates
the moments of $X_{n}$:%
\begin{equation*}
\frac{\partial ^{p}\Theta _{n}}{\partial z^{p}}(\beta ,0)=\frac{\displaystyle%
\int d\sigma _{n}(\mathbf{x};\sqrt{n})~X_{n}^{p}\exp \left\{ -\dfrac{\beta }{%
2}\left( \mathbf{x},J\mathbf{x}\right) \right\} }{\displaystyle\int d\sigma
_{n}(\mathbf{x};\sqrt{n})~\exp \left\{ -\dfrac{\beta }{2}\left( \mathbf{x},J%
\mathbf{x}\right) \right\} }\equiv \left\langle X_{n}^{p}\right\rangle
\end{equation*}%
and the same procedure of the previous subsection can be used to evaluate
the moment generating function $\Theta _{n}$. Despite of fact that $\mathbf{h%
}$ in this case goes to $\mathbf{0}$ when $n\rightarrow \infty $, the ratio (%
\ref{thetan}) converges to a nontrivial ($\neq 1$) function of $z$ as we
shall see in the following.

Repeating (\ref{In}) - (\ref{I2}) with $I_{n}$ replaced by an auxiliary
function for (\ref{Qnzeta}) yields%
\begin{eqnarray}
K_{n} &=&\frac{1}{S_{n}}\int_{\mathbb{R}^{n}}\prod_{i=1}^{n}dx_{i}~\exp
\left\{ \frac{-1}{2}\left( \mathbf{x},\left( J-\kappa \right) \mathbf{x}%
\right) +\left( \mathbf{h},\mathbf{x}\right) \right\}  \notag \\
&=&\frac{2^{n/2-1}~\Gamma (n/2)}{n^{(n-1)/2}\sqrt{\det \left( J-\kappa
\right) }}\exp \left\{ \frac{1}{2}\left( \mathbf{h},\frac{1}{J-\kappa I}%
\mathbf{h}\right) \right\}  \notag \\
&=&\sqrt{n}\int_{0}^{\infty }\frac{ds}{s}~\exp \left\{ n~g_{n}(s)\right\}
\label{Kn}
\end{eqnarray}%
with%
\begin{equation*}
g_{n}(s)=\frac{\kappa }{2}s^{2}+\ln s+f_{n}(s^{2},s)
\end{equation*}%
where the free energy

\begin{equation*}
f_{n}\left( \beta ,z\right) =\frac{1}{n}\ln Q_{n}\left( \beta ,z;\mathbf{h}%
,J\right) ~
\end{equation*}%
is a smooth function of $(\beta ,z)$ such that $f_{n}\left( s^{2},0\right)
=f_{n}\left( s^{2}\right) $ is given by (\ref{fn}).

We continue from the last two equations of (\ref{Kn}):%
\begin{eqnarray}
\lim_{n\rightarrow \infty }\frac{1}{n}\ln K_{n} &=&-\frac{1}{2}-\frac{1}{2}%
\mathbb{E}\ln \left( J-\kappa \right) +\frac{1}{2}\mathbb{E}_{\mathbf{h}%
}\left( J-\kappa \right) ^{-1}  \notag \\
&=&\frac{\kappa }{2}\bar{s}^{2}+\ln \bar{s}+f(\bar{s}^{2},\bar{s})
\label{lnKn}
\end{eqnarray}%
where $\bar{s}=\bar{s}(\kappa )$ is the solution of equation (\ref{nu2})
with $\mu $ and $f^{\prime }(\bar{s}^{2})$ replaced by $\kappa $ and the
derivative of $f(s^{2},s)$ with respect to $s^{2}$ (recall $f(s^{2},s)$ is
an even function of $s$):%
\begin{equation}
\bar{s}^{2}=-\left( \kappa +2\left. \frac{d}{ds^{2}}f(s^{2},s)\right\vert
_{s=\bar{s}}\right) ^{-1}  \label{sbar}
\end{equation}
and, if $P_{\mathbf{h}}=\mathbf{hh}^{T}/\left\Vert \mathbf{h}\right\Vert
^{2} $ denotes the projector in the $\mathbf{h}$ direction and $\left\{
\lambda _{j},E_{j}\right\} _{j=1}^{n}$ are the spectral elements of $J$: $J=%
\displaystyle\sum_{j=1}^{n}\lambda _{j}~E_{j}$, we have%
\begin{equation}
\mathbb{E}_{\mathbf{h}}\left( J-\kappa \right) ^{-1}=\lim_{n\rightarrow
\infty }\frac{\left\Vert \mathbf{h}\right\Vert ^{2}}{n}\text{Tr}\frac{1}{%
J-\kappa }P_{\mathbf{h}}=\lim_{n\rightarrow \infty }\frac{1}{n}\sum_{j=1}^{n}%
\frac{1}{\lambda _{j}-\kappa }\left\Vert E_{j}\mathbf{h}\right\Vert ^{2}~.
\label{expect}
\end{equation}

When $J=-\Delta $, Fourier analysis gives 
\begin{equation*}
\mathbb{E}_{\mathbf{h}}\left( -\Delta -\kappa \right) ^{-1}=\frac{\beta }{%
\left( 2\pi \right) ^{d}}\int_{[-\pi ,\pi ]^{d}}d^{d}k\frac{\left\vert \hat{h%
}(\mathbf{k})\right\vert ^{2}}{\omega (k)-\kappa }
\end{equation*}%
with%
\begin{equation*}
\hat{h}(k)=\sum_{x\in \mathbb{Z}^{d}}h_{x}~\exp \left( ik\cdot x\right) ~.
\end{equation*}%
Since $\mathbf{1}=\left( 1/\sqrt{n}\right) \left( 1,\ldots ,1\right) $ is
the unique eigenvector of (\ref{laplacean}) associated with the eigenvalue $%
0 $, only the zero--mode contributes to the expectation (\ref{expect}) with $%
\mathbf{h}=\mathbf{1}$:%
\begin{equation}
\mathbb{E}_{\mathbf{h}}\left( -\Delta -\kappa \right) ^{-1}=\left( \mathbf{h}%
,\left( -\Delta -\kappa \right) ^{-1}\mathbf{h}\right) =\frac{-z^{2}}{%
n\kappa }~.  \label{Eh}
\end{equation}%
Assuming $\mathbf{1}$ an eigenvector of $J$ (orthogonal to the complementary
space in view of $J=J^{T}$) with associate eigenvalue $\lambda =0$ the same
result holds with $-\Delta $ replaced by $J$. We shall continue our
calculation of the generating function $\Theta _{n}$ with $J$ satisfying the
assumptions of admissible coupling matrices.

The free energy function is obtained by equating the two lines of (\ref{lnKn}%
) up to order $1/n$ and proceeding as in equations (\ref{nu2})--(\ref{f}): 
\begin{equation}
nf_{n}\left( \beta ,z\right) =-\frac{n}{2}\left( \beta \kappa +\ln \left(
e\beta \right) +\mathbb{E}\ln \left( J-\kappa \right) \right) -\frac{z^{2}}{%
2\kappa }+O(1)~,  \label{fn1}
\end{equation}%
where $\kappa _{n}=\kappa _{n}\left( \beta ,z\right) $ is the solution of%
\begin{equation}
\beta =~\mathbb{E}\frac{1}{-\Delta -\kappa }+\frac{z^{2}}{n\kappa ^{2}}~,
\label{beta}
\end{equation}%
and the order $1$ term is independent of $z$ by (\ref{Itilde}) for $K_{n}$
together with (\ref{kappa}) below.

Denoting by $w_{n}(\kappa ,z)$ the r.h.s. of (\ref{beta}), $w_{n}$ is a
decreasing function of $\kappa $ for $\kappa <0$ with $w_{n}(-\infty ,z)=0$
and $\lim_{\kappa \rightarrow 0}w_{n}(\kappa ,z)=\infty $. By the implicit
function theorem, the solution $\kappa _{n}\left( \beta ,z\right) $ of (\ref%
{beta}) is the unique real analytic function of $z^{2}$ in a neighborhood of 
$z^{2}=0$. Consequently, $\kappa _{n}\left( \beta ,0\right) =\mu \left(
\beta \right) $ for every $n$, $\lim_{n\rightarrow \infty }\kappa _{n}\left(
\beta ,z\right) =\mu \left( \beta \right) $ uniformly in $z$, where $\mu
(\beta )$ is the solution of (\ref{nu}) and 
\begin{equation}
\kappa =\mu +\frac{c}{n}z^{2}+o\left( \frac{z^{2}}{n}\right) ~  \label{kappa}
\end{equation}%
by Taylor theorem, where%
\begin{equation*}
c=\left[ \mathbb{E}\left( 1-J/\mu \right) ^{-2}\right] ^{-1}~
\end{equation*}%
is obtained by plugging the resolvent equation%
\begin{equation*}
\frac{1}{J-\kappa }-\frac{1}{J-\mu }=\left( \kappa -\mu \right) \frac{1}{%
J-\kappa }\frac{1}{J-\mu }
\end{equation*}%
into (\ref{beta}) together with (\ref{nu}) and (\ref{kappa}).

\medskip

\noindent \textit{Proof of Theorem \ref{CLT}. }Substituting $\exp \left(
nf_{n}(\beta ,z)\right) $ in the numerator and denominator (with $z=0$) of (%
\ref{thetan}), taking into account that $O(1)$ term in (\ref{fn1}) does not
depend on $z$, gives%
\begin{equation}
\Theta _{n}(\beta ,z)=\exp \left( -n\frac{\beta }{2}\left( \kappa -\mu
\right) -\frac{n}{2}\mathbb{E}\ln \frac{J-\kappa }{J-\mu }-\frac{z^{2}}{%
2\kappa }\right) \left( 1+o\left( \frac{1}{n}\right) \right)  \label{FnFn}
\end{equation}%
and it suffices to verify that 
\begin{equation}
\ln \Theta (\beta ,z)=\lim_{n\rightarrow \infty }\ln \Theta _{n}(\beta ,z)
\label{FF}
\end{equation}%
converges uniformly in compacts of $z$ and for every $\beta <\beta _{c}$ to
a limit proportional to $z^{2}$. The limit (\ref{FF}) exists by the same
reasons employed to show existence of $\lim_{n\rightarrow \infty
}f_{n}(s^{2})$. Plugging (\ref{kappa}) into (\ref{FnFn}) together (\ref{nu})
gives%
\begin{equation*}
\ln \Theta (\beta ,z)=\frac{-z^{2}}{2}\left( c\beta -c\mathbb{E}\frac{1}{%
J-\mu }+\frac{1}{\mu }\right) =\frac{-z^{2}}{2\mu (\beta )}~.
\end{equation*}

\hfill $\Box $

\section{Hierarchical Spherical Model\label{HSM}}

\setcounter{equation}{0} \setcounter{theorem}{0}

\subsection{Hierarchical Laplacean}

Paiva and Perez \cite{PP} have investigated the semi--groups generated by $d$%
--dimensional Hierarchical Laplacean $-\Delta $ in the presence of disorder
by using, for the first time, spectral analysis. Although $-\Delta $ has
discrete spectrum, they have shown that $\exp \left( t\Delta \right) \delta
_{0}$, with $\delta _{0}$ localized at origin, diffuses. We quote \cite{Kr}
and references therein for spectral localization in hierarchical Anderson
model. Here, in order to apply the limit theorem established in Subsection %
\ref{SMG} we need the spectral theorem for homogeneous hierarchical
matrices. We extend the work of Watanabe \cite{W} to arbitrary spectral
dimension $d$ (see Remark \ref{specdim}).

Given integer numbers $L$, $K>1$ and $d\geq 1$, let%
\begin{equation*}
\Lambda _{K}=\left\{ 0,1,\ldots ,L^{K}-1\right\} ^{d}\subset \mathbb{Z}^{d}
\end{equation*}%
be a hypercube with cardinality $\left\vert \Lambda _{K}\right\vert
=L^{dK}=n $. Let $\mathbf{\theta }=\left( \theta _{1},\ldots ,\theta
_{K}\right) $ denote the coordinates of a point $i\in \Lambda _{K}$ written
in the $L^{d}$ base%
\begin{equation*}
i=\sum_{k=1}^{K}\theta _{k}L^{k-1}\ ,\ \ \theta _{k}\in \left\{ 0,1,\ldots
,L-1\right\} ^{d}~.
\end{equation*}%
From now on we use these coordinates to index components of a vector $%
u=\left( u_{\mathbf{\theta }}\right) $ in $\mathbb{R}^{\Lambda _{K}}$.

Let $B:\mathbb{R}^{\Lambda _{K}}\longrightarrow \mathbb{R}^{\Lambda _{K-1}}$
be the block operator%
\begin{equation}
\left( Bu\right) _{\mathbf{\tau }}=\frac{1}{L^{d/2}}\sum_{\theta \in \left\{
0,1,\ldots ,L^{K}-1\right\} ^{d}}u_{\left( \theta ,\mathbf{\tau }\right) }
\label{B}
\end{equation}%
and let $B^{\ast }:\mathbb{R}^{\Lambda _{K-1}}\longrightarrow \mathbb{R}%
^{\Lambda _{K}}$ be its adjoint%
\begin{equation*}
\left( B^{\ast }v,u\right) _{\Lambda _{K}}=\left( v,Bu\right) _{\Lambda
_{K-1}}
\end{equation*}%
with respect to the inner product $\left( u,w\right) _{\Lambda _{K}}=%
\displaystyle\sum_{\mathbf{\theta }}u_{\mathbf{\theta }}~w_{\mathbf{\theta }%
} $:%
\begin{equation*}
\left( B^{\ast }v\right) _{(\theta ,\mathbf{\tau })}=\frac{1}{L^{d/2}}v_{%
\mathbf{\tau }}~.
\end{equation*}

Define a real symmetric matrix in $\mathbb{R}^{n}$%
\begin{equation}
J=\sum_{k=1}^{K}L^{-2k}\left( B^{\ast }\right) ^{k}B^{k}~~.  \label{J}
\end{equation}

\begin{proposition}
\label{dyson}The associate quadratic form of $J$ for $L^{d}=2$ and $%
d=2/\left( \alpha -1\right) $ gives the hierarchical energy%
\begin{eqnarray*}
-H &=&\sum_{k=1}^{K}2^{-\alpha k}\sum_{r=1}^{2^{K-k}}\left( S_{k,r}\right)
^{2} \\
S_{k,r} &=&\sum_{(r-1)2^{k}<j\leq r2^{k}}\sigma _{j}
\end{eqnarray*}%
introduced by Dyson \cite{D} in his study of the Ising model with $%
1/\left\vert i-j\right\vert ^{\alpha }$--interaction.
\end{proposition}

\begin{remark}
Note that $\alpha (d)=\left( d+2\right) /d$ ranges from $2$ to $1$ as $d$
varies from $2$ to $\infty $.
\end{remark}

\noindent \textit{Proof. }Setting $L^{d}=2$, we have $S_{k,r}=2^{k/2}\left(
B^{k}\sigma \right) _{\mathbf{\theta }}$ for some $\mathbf{\theta }\in
\Lambda _{K-k}$ and 
\begin{equation*}
-H=\sum_{k=1}^{K}2^{-\alpha k+k}\left( B^{k}\sigma ,B^{k}\sigma \right)
_{\Lambda _{K-k}}=\sum_{k=1}^{K}L^{-2k}\left( \sigma ,\left( B^{\ast
}\right) ^{k}B^{k}\sigma \right) _{\Lambda _{K}}=\left( \sigma ,J\sigma
\right) _{\Lambda _{K}}
\end{equation*}%
as claimed.

\hfill $\Box $

We require the hierarchical Laplacean $-\Delta $ satisfies 
\begin{equation}
-\Delta \mathbf{1}=0~  \label{Delta}
\end{equation}%
(Theorem \ref{TS} below shows that $0$ is also a simple eigenvalue). Writing%
\begin{equation*}
J=\Delta +\mu _{0}I
\end{equation*}%
together with (\ref{Delta}), (\ref{J}) and (\ref{B}), we have 
\begin{equation}
\mu _{0}=\left( \mathbf{1},J\mathbf{1}\right) _{\Lambda _{K}}=\frac{1}{L^{dK}%
}\sum_{k=1}^{K}L^{-2k}\left( B^{k}\mathbf{1},B^{k}\mathbf{1}\right)
_{\Lambda _{K-k}}=\sum_{k=1}^{K}L^{-2k}  \label{mu0}
\end{equation}%
and%
\begin{equation}
-\Delta =\sum_{k=1}^{K}L^{-2k}\left( -\left( B^{\ast }\right)
^{k}B^{k}+I\right) ~  \label{-D}
\end{equation}%
generates a stochastic semi--group.

Now we observe that 
\begin{equation}
B^{k}\left( B^{\ast }\right) ^{k}=BB^{\ast }=I  \label{BB}
\end{equation}%
holds for every $k=1,\ldots ,K$ and%
\begin{equation}
P_{k}=\left( B^{\ast }\right) ^{k}B^{k}  \label{PBB}
\end{equation}%
is an orthogonal $P_{k}=P_{k}^{\ast }$ projection matrix $P_{k}^{2}=P_{k}$
on the subspace of vectors in $\mathbb{R}^{\Lambda _{K}}$ which assumes
constant value over blocks of size $L^{dk}$. It follows from (\ref{BB})%
\begin{eqnarray}
P_{j}P_{k} &=&\left( B^{\ast }\right) ^{j}B^{j}\left( B^{\ast }\right)
^{k}B^{k}=\left( B^{\ast }\right) ^{j}B^{j-k}B^{k}=P_{j}  \notag \\
P_{k}P_{j} &=&\left( B^{\ast }\right) ^{k}B^{k}\left( B^{\ast }\right)
^{j}B^{j}=\left( B^{\ast }\right) ^{k}\left( B^{\ast }\right)
^{j-k}B^{j}=P_{j}  \label{P}
\end{eqnarray}%
hold for any $j>k$ and we have the following inclusions%
\begin{equation}
P_{K}<P_{K-1}<\cdots <P_{1}<P_{0}\equiv I~  \label{incl}
\end{equation}%
in the sense that $A<B$ if, and only if, $\left( u,Au\right) _{\Lambda
_{k}}<\left( u,Bu\right) _{\Lambda _{k}}$ holds for all $u\in \mathbb{R}%
^{\Lambda _{K}}$.

Let%
\begin{equation}
Q_{k}=P_{k}-P_{k+1}  \label{Q}
\end{equation}%
for $k=0,1,\ldots ,K-1$ and%
\begin{equation*}
Q_{K}=P_{K}~
\end{equation*}%
be the block fluctuation operator.

\begin{theorem}[Spectral]
\label{TS}The collection $\left\{ Q_{k}\right\} _{k=0}^{K}$ of $n\times n$
real orthogonal projection matrices%
\begin{equation}
Q_{j}Q_{k}=\delta _{jk}Q_{k}  \label{QQQ}
\end{equation}%
are the spectral partition of unit 
\begin{equation*}
I=\sum_{k=0}^{K}Q_{k}
\end{equation*}%
and 
\begin{equation}
f\left( -\Delta \right) =\sum_{k=0}^{K}f\left( \lambda _{k}\right) ~Q_{k}
\label{Dl}
\end{equation}%
holds with%
\begin{equation}
\lambda _{k}=\frac{L^{-2k}-L^{-2K}}{L^{2}-1}~  \label{l}
\end{equation}%
for any continuous function $f:\left[ 0,1/(L^{2}-1)\right] \longrightarrow 
\mathbb{R}$. It follows that $-\Delta $ is a positive definite matrix where $%
\lambda _{k}$, $k=0,\ldots ,K-1$, is an eigenvalue of multiplicity $%
L^{d(K-k)}(1-L^{-d})$ and $\lambda _{K}=0$ a simple eigenvalue.
\end{theorem}

\noindent \textit{Proof.} The prove is essentially given in \cite{W}. By (%
\ref{Q}) and (\ref{P})%
\begin{eqnarray*}
Q_{j}Q_{k} &=&\left( P_{j}-P_{j+1}\right) \left( P_{k}-P_{k+1}\right) \\
&=&P_{j}\left( P_{k}-P_{k+1}\right) -P_{j+1}\left( P_{k}-P_{k+1}\right) \\
&=&\left( P_{j}-P_{j}\right) -\left( P_{j+1}-P_{j+1}\right) =0
\end{eqnarray*}%
for any $k<j<K$ and the same holds for $j<k<K$. For $j<k=K$,%
\begin{equation*}
Q_{j}Q_{K}=\left( P_{j}-P_{j+1}\right) P_{K}=P_{K}-P_{K}=0
\end{equation*}%
and for $j=k$%
\begin{equation*}
Q_{k}Q_{k}=\left( P_{k}-P_{k+1}\right) \left( P_{k}-P_{k+1}\right)
=P_{k}+P_{k+1}-2P_{k+1}=Q_{k}~.
\end{equation*}%
By definition, 
\begin{equation*}
\sum_{k=0}^{K}Q_{k}=\sum_{k=0}^{K-1}\left( P_{k}-P_{k+1}\right)
+P_{K}=P_{0}-P_{K}+P_{K}=I~.
\end{equation*}%
Finally, by (\ref{-D}), (\ref{PBB}) and (\ref{Q}), we have%
\begin{eqnarray}
-\Delta &=&\sum_{j=1}^{K}L^{-2j}\left( -P_{j}+I\right)  \notag \\
&=&\sum_{j=1}^{K}L^{-2j}\sum_{k=0}^{j-1}Q_{k}  \notag \\
&=&\sum_{k=0}^{K-1}\left( \sum_{j=k+1}^{K}L^{-2j}\right) Q_{k}+0\cdot Q_{K}
\label{hL}
\end{eqnarray}%
which gives (\ref{Dl}) with $f(x)=x$. It follows by (\ref{QQQ}) that (\ref%
{Dl}) holds for any polynomial and, by Weierstrass approximation theorem,
for any uniformly continuous function.

Since $P_{k}$ projects on vectors in $\Lambda _{K}$ which are constant over
disjoint blocks of size $L^{dk}$, the rank of $P_{k}$ is 
\begin{equation*}
\text{rank}P_{k}=L^{d(K-k)}~.
\end{equation*}%
By definition (\ref{Q}) together with the inclusions (\ref{incl}), the rank
of the block fluctuation projector $Q_{k}$ is%
\begin{equation}
\text{rank}Q_{k}=L^{d(K-k)}-L^{d(K-k-1)}  \label{rank}
\end{equation}%
for $k=1,\ldots ,K-1$ and 
\begin{equation*}
\text{rank}Q_{K}=1~
\end{equation*}%
and these concludes the prove of Theorem \ref{TS}.

\hfill $\Box $

\begin{remark}
\label{specdim}The spectral measure $\mu ^{K}$ associated with the vector $%
\delta _{\mathbf{\theta }}=\left( \delta _{\mathbf{\theta },j}\right) _{j\in
\Lambda _{K}}$ defined by%
\begin{equation*}
\left( \delta _{\mathbf{\theta }},f\left( -\Delta \right) \delta _{\mathbf{%
\theta }}\right) =\int_{-\infty }^{\infty }f(x)~d\mu _{\theta }^{K}\left(
x\right)
\end{equation*}%
for every bounded Borel function $f:\mathbb{R}\longrightarrow \mathbb{C}$,
is given by%
\begin{equation*}
d\mu ^{K}\left( x\right) =\sum_{k=0}^{K-1}\frac{L^{d}-1}{L^{d(k+1)}}\delta
\left( x-\lambda _{k}\right) ~dx+\frac{1}{L^{dK}}\delta \left( x\right) ~dx~,
\end{equation*}%
by inspection of the matrix elements $Q_{\mathbf{\theta \theta }^{\prime }}$%
, and is independent of $\mathbf{\theta }$. As $n=L^{dK}$ tends to infinity, 
$\mu ^{\infty }$ is the unique weak--$\ast $ limit point of the
corresponding empirical measure (\ref{rhon}) (see Theorem $1.2$ of
Kritchevski \cite{Kr}). The number%
\begin{equation*}
d:=2\lim_{t\downarrow 0}\frac{\ln \mu ^{\infty }\left( \left[ 0,t\right]
\right) }{\ln t}
\end{equation*}%
is called spectral dimension of $-\Delta $.
\end{remark}

\subsection{The Free Energy}

To compute the free energy (\ref{f}) of the spherical model associated with $%
\beta >0$ and hierarchical Laplacean\ matrix $J=-\Delta $ we need to
evaluate the expectation $\mathbb{E}$ with respect to the empirical measure
of eigenvalues of $-\Delta $ in both the last term of the r.h.s. of (\ref{f}%
) and in the implicit equation (\ref{nu}) for $\mu =\mu (\beta )$.

Using Theorem \ref{TS} together with (\ref{T}) and linearity of trace, we
have 
\begin{eqnarray*}
\frac{1}{L^{dK}}\text{Tr}f(-\Delta ) &=&\frac{1}{L^{dK}}\sum_{k=0}^{K}f(%
\lambda _{k})~\text{Tr}Q_{k} \\
&=&\left( 1-L^{-d}\right) \sum_{k=0}^{K-1}L^{-dk}~f(\lambda _{k})+\frac{1}{%
L^{dK}}~f(\lambda _{K})
\end{eqnarray*}%
in view of the fact that the eigenvalues of $Q_{k}$ are $0$ and $1$ together
with (\ref{rank}). Hence, the subsequence $\left\{ f_{n_{K}}(s^{2})\right\}
_{K\in \mathbb{N}}$ of the free energy with $n_{K}=L^{dK}$ converges to (\ref%
{f}) where%
\begin{equation}
\mathbb{E}\ln \left( -\Delta -\mu \right) =\left( 1-L^{-d}\right)
\sum_{k=0}^{\infty }L^{-dk}\ln \left( \frac{L^{-2k}}{L^{2}-1}-\mu \right)
\label{Tln}
\end{equation}%
and $\mu =\mu (\beta )$ solves%
\begin{eqnarray}
\beta &=&\mathbb{E}\left( -\Delta -\mu I\right) ^{-1}  \notag \\
&=&\left( 1-L^{-d}\right) \sum_{k=0}^{\infty }L^{-dk}\frac{L^{2}-1}{%
L^{-2k}-\mu (L^{2}-1)}+\rho _{0}  \label{TD}
\end{eqnarray}%
including the $0$--eigenvalue contribution $\rho _{0}=\mathbb{E}\mathcal{P}%
_{0}\left( -\Delta -\mu I\right) ^{-1}$ which, by Remark \ref{condensate},
may have macroscopic occupation.

Analogously to (\ref{0}), we have%
\begin{eqnarray*}
\rho _{0} &\geq &\beta -\left( 1-L^{-d}\right) \left( L^{2}-1\right)
\sum_{k=0}^{\infty }L^{-(d-2)k} \\
&=&\beta -\frac{\left( 1-L^{-d}\right) \left( L^{2}-1\right) }{1-L^{-d+2}}
\end{eqnarray*}%
which is strictly positive provided $d>2$ and $\beta >\beta _{c}(d,L)$ where 
\begin{equation*}
\beta _{c}(d,L)=\frac{\left( 1-L^{-d}\right) \left( L^{2}-1\right) }{%
1-L^{-d+2}}~
\end{equation*}%
is the critical inverse temperature of the hierarchical spherical model.

\begin{remark}
The geometric mul\-ti\-pli\-ci\-ty $(L^{d}-1)L^{d(K-k-1)}$ of each
eigenvalue $\lambda _{k}$, $k=0,\ldots ,K-1$, of the hierarchical Laplacean (%
\ref{hL}) can be lift by fluctuation projectors $Q_{k,\mathbf{\theta }}$
depending on the index $\mathbf{\theta }\in \Lambda _{K-k}$, such that 
\begin{equation*}
Q_{k,\mathbf{\theta }}Q_{k,\mathbf{\theta }^{\prime }}=\delta _{\mathbf{%
\theta \theta }^{\prime }}Q_{k,\mathbf{\theta }}~,
\end{equation*}%
and a nonhomogeneous Laplacean can be defined by%
\begin{equation*}
-\Delta ^{\mathrm{nh}}=\sum_{k=0}^{K}\sum_{\mathbf{\theta }\in \Lambda
_{K-k}}\lambda _{k,\mathbf{\theta }}~Q_{k,\mathbf{\theta }}
\end{equation*}%
with%
\begin{equation*}
\lambda _{k,\mathbf{\theta }}=c~\lambda _{k}~\exp \left\{ X_{k,\mathbf{%
\theta }}\right\}
\end{equation*}%
where $\lambda _{k}$ is given by (\ref{l}), $\left\{ X_{k,\mathbf{\theta }%
}\right\} $ chosen according to a common probability distribution $\mathbb{P}
$ with mean $0$ with $c^{-1}=\mathbb{E}\exp \left( X_{k,\mathbf{\theta }%
}\right) <\infty $. In this case, we have%
\begin{eqnarray*}
\frac{1}{L^{dK}}\text{Tr}f(-\Delta ) &=&\left( 1-L^{-d}\right)
\sum_{k=0}^{K-1}L^{-dk}~\frac{1}{L^{d(K-k)}}\sum_{\mathbf{\theta }\in
\Lambda _{K-k}}f(c\lambda _{k}\exp \left\{ X_{k,\mathbf{\theta }}\right\} )+%
\frac{1}{L^{dK}}~f(\lambda _{K}) \\
&\longrightarrow &\left( 1-L^{-d}\right) \sum_{k=0}^{\infty }L^{-dk}\mathbb{E%
}f(c\lambda _{k}\exp \left\{ X_{k,\mathbf{0}}\right\} )=\mathbb{E}f(-\Delta )
\end{eqnarray*}%
for almost every $\left\{ X_{k,\mathbf{0}}\right\} $ with respect to
distribution $\mathbb{P}$, by the law of large numbers. Here $\mathbb{E}f$
denotes the expectation with respect to the product measure $d\rho (\lambda )%
\mathbb{P}\left( dx\right) $ with $\rho $ the empirical distribution
relative to the eigenvalues $\left\{ \lambda _{k}\right\} $ and $\mathbb{P}$
the common distribution of variables $X_{k,\mathbf{\theta }}$. Expressions (%
\ref{f}), (\ref{Tln}) and (\ref{TD}) are obtained accordingly. For instance,%
\begin{equation*}
\mathbb{E}\left( -\Delta -\mu I\right) ^{-1}=\left( 1-L^{-d}\right)
\sum_{k=0}^{\infty }L^{-dk}\mathbb{E}\frac{L^{2}-1}{cL^{-2k}\exp \left\{
X_{k,\mathbf{0}}\right\} -\mu (L^{2}-1)}+\rho _{0}~.
\end{equation*}
\end{remark}

\subsection{Continuum Hierarchical Laplacean\label{CHL}}

Hierarchical Laplaceans have discrete eigenvalues. We shall now consider a
continuous version obtained by a limit procedure.

We shall take $L\downarrow 1$ simultaneously to $K\rightarrow \infty $
maintaining $K\ln L$ fixed equal to $C\in \mathbb{R}_{+}\cup \left\{ \infty
\right\} $. Equation (\ref{TD}), for instance, reads 
\begin{eqnarray}
\mathbb{E}\left( -\Delta -\mu I\right) ^{-1} &=&\lim_{L\downarrow 1}\left(
1-L^{-d}\right) \sum_{k=0}^{K(L)}L^{-dk}\left( \frac{L-1}{\left(
L^{2}-1\right) }(L^{-2k}-L^{-2K})-\mu \right) ^{-1}  \notag \\
&=&d\int_{0}^{C}\frac{2}{\exp \left( -2y\right) -\exp \left( -2C\right)
-2\mu }e^{-dy}dy  \notag \\
&=&~\int_{0}^{(1-\exp \left( -2C\right) )/2}\frac{1}{\lambda -\mu }d\rho
(\lambda )  \label{Delta-mu}
\end{eqnarray}%
where $d\rho (\lambda )=\rho ^{\prime }(\lambda )d\lambda $ is absolutely
continuous with respect to the Lebesgue measure $d\lambda $ with 
\begin{equation*}
\rho ^{\prime }(\lambda )=2^{d/2}\frac{d}{2}\left( \lambda +\frac{\exp
\left( -2C\right) }{2}\right) ^{d/2-1}~
\end{equation*}%
if $\lambda \in \left[ 0,(1-\exp \left( -2C\right) )/2\right] $ and $0$
otherwise. So, the empirical measure for the eigenvalues of the hierarchical
Laplacean $-\Delta $ with $L\downarrow 1$ converges provided $K=K(L)$
increases faster than $C\left( \ln L\right) ^{-1}$. We take $C=\infty $, for
simplicity.

Accordingly, equations (\ref{f}), (\ref{Tln}) and (\ref{TD}) holds with $%
\lambda _{k}$ replaced by $\lambda _{k}\left( L-1\right) $ and%
\begin{eqnarray*}
\rho _{0} &\geq &\beta -\lim_{L\downarrow 1}\frac{\left( 1-L^{-d}\right)
\left( L^{2}-1\right) }{\left( 1-L^{-d+2}\right) \left( L-1\right) } \\
&=&\beta -\frac{2d}{d-2}
\end{eqnarray*}%
has a strictly positive limit provided $d>2$ and $\beta >\beta _{c}(d)$
where 
\begin{equation}
\beta _{c}(d)=\frac{2d}{d-2}~.  \label{betac}
\end{equation}%
Note that 
\begin{equation*}
\mathbb{E}\left( -\Delta \right) ^{-1}=\int_{0}^{1/2}\lambda ^{-1}d\rho
(\lambda )=2^{d/2}\frac{d}{2}\int_{0}^{1/2}\lambda ^{d/2-2}d\lambda =\frac{2d%
}{d-2}
\end{equation*}%
and a phase transition of Bose--Einstein condensation type occur at the
critical inverse temperature $\beta _{c}=\mathbb{E}\left( -\Delta \right)
^{-1}$ as in the spherical model with the usual Laplacean interaction.

We now compute integral (\ref{Delta-mu}) with $C=\infty $ and $d=4$, for
comparison purposes. For $\mu \leq 0$, we continue%
\begin{equation*}
\mathbb{E}\left( -\Delta -\mu I\right) ^{-1}=8~\int_{0}^{1/2}\frac{\lambda }{%
\lambda -\mu }d\lambda =4\left( 2\mu \ln \left( 1-\frac{1}{2\mu }\right)
+1\right)
\end{equation*}%
and equation (\ref{nu}) with $\Delta $ given by the hierarchical Laplacean
at $d=4$ and $\bar{s}^{2}=\beta $, reads%
\begin{equation}
1-\frac{\beta }{4}=-2\mu \ln \left( 1-\frac{1}{2\mu }\right) ~.
\label{beta-mu}
\end{equation}

\section{Convergence to Spherical Model\label{CSM}}

\setcounter{equation}{0} \setcounter{theorem}{0}

\subsection{The $O(N)$ Heisenberg Model}

The partition function of the $O\left( N\right) $ symmetric Heisenberg model
in a $d$--dimensional cubic box $\Lambda \subset \mathbb{Z}^{d}$ of
cardinality $n=L^{d}$ is given by%
\begin{equation}
Z_{n}^{(N)}(\beta ,A)=\frac{1}{S_{N}^{n}}\int_{\mathbb{R}^{nN}}\exp \left\{ -%
\frac{\beta }{2}\left( \mathbf{x},A\mathbf{x}\right) \right\}
\prod_{j=1}^{n}d\sigma _{0}^{(N)}\left( x_{j}\right)   \label{Zn}
\end{equation}%
where $\mathbf{x}=\left( x_{1},\ldots ,x_{n}\right) $ is a $n$--tuple with
each $x_{j}\in \mathbb{R}^{N}$, $A=J\otimes I$ is the tensor product of a $%
n\times n$ coupling matrix $J$ with the $N\times N$ identity matrix $I$ and $%
\sigma _{0}^{(N)}\left( dx\right) \equiv \left. \sigma _{N}\left( dx;\sqrt{N}%
\right) \right/ S_{N}$ is the \textquotedblleft a priori\textquotedblright\
uniform probability measure on the $N$--dimensional sphere $\left\vert
x\right\vert ^{2}=N$ of radius $\sqrt{N}$ with surface area $S_{N}$ given by
(\ref{Sn}). The inner product on $\mathbb{R}^{n}\otimes \mathbb{R}^{N}$ is
denoted here by $\left( \mathbf{x},\mathbf{y}\right) =\displaystyle%
\sum_{i=1}^{n}x_{i}\cdot y_{i}$.

The expected value with respect to the Heisenberg measure $\nu _{n}^{(N)}$
(see (\ref{equ})) is defined by%
\begin{equation*}
\langle F\rangle _{\nu _{n}^{(N)}}=\frac{1}{S_{N}^{n}Z_{n}^{(N)}}\int_{%
\mathbb{R}^{nN}}F(\mathbf{x})\exp \left\{ -\frac{\beta }{2}\left( \mathbf{x}%
,A\mathbf{x}\right) \right\} \prod_{j=1}^{n}d\sigma _{0}^{(N)}\left(
x_{j}\right) 
\end{equation*}%
and from here on, $A$ in $\langle \cdot \rangle _{\nu _{n}^{(N)}}=\langle
\cdot \rangle _{\nu _{n}^{(N)}}(\beta ,A)$ is assumed to be a sequence of
admissible coupling matrices in the following sense.

\begin{definition}
\label{adm1}A sequence $A=\left\{ A_{n}\right\} _{n\geq 1}$ of coupling
matrices is an admissible reflection positive sequence if each $A_{n}$ is
admissible in the sense of Definition \ref{adm} and $\langle \cdot \rangle
_{\nu _{n}^{(N)}}(\beta ,A)$ satisfies the \textit{reflection positivity}
condition (\ref{pos}).
\end{definition}

Kunz and Zumbach \cite{KZ} have devised a way of proving convergence of the
free energy of the Heisenberg model to the spherical model for nearest
neighbor interactions. We shall first show that their method holds for
admissible reflection positive coupling matrices and prove Theorem \textit{%
\ref{convergence1}} afterwards.

\begin{theorem}
\label{convergence}The finite volume free energy of the Heisenberg model%
\begin{equation}
f_{n}^{(N)}(\beta )=\frac{1}{n\cdot N}\ln Z_{n}^{(N)}(\beta ,A)~  \label{fnN}
\end{equation}%
with admissible reflection positive sequence of coupling matrices $A$,
converges 
\begin{equation}
\lim_{n,N\rightarrow \infty }f_{n}^{(N)}(\beta )=f(\beta )  \label{ff}
\end{equation}%
to the spherical model free energy (\ref{f}) as $n$, $N$ goes to infinity in
any order.
\end{theorem}

The original proof by Kac and Thompson \cite{KT} asserts that (\ref{ff})
holds for all coupling matrix $J$ satisfying translation invariance $%
J_{ij}=g\left( i-j\right) $. It turns out that their proof has a serious gap
fixed in \cite{KZ} only for the usual Laplacean given by (\ref{laplacean}).
Our presentation, based on an unpublished Appendix of \cite{KZ}, uses the
Laplace method discussed in Subsection \ref{SFE} and is written for
admissible coupling matrices. Consequently, it works for the hierarchical
Laplacean matrix coupling (\ref{J}) as well.

Reflection Positivity is the missing ingredient. For a basic exposition of
the abstract theory see \cite{FILS}. If $P$ denotes a plane perpendicular to
coordinate axes which divides $\Lambda $ into two halves $\Lambda _{\pm }$: 
\begin{equation*}
\Lambda =\Lambda _{+}\cup \Lambda _{-}\qquad \mathrm{and}\ \qquad \Lambda
_{+}\cap \Lambda _{-}=\varnothing
\end{equation*}%
($P$ cuts bonds perpendicularly and do not intercept sites of $\Lambda $),
let 
\begin{equation*}
r:\Lambda _{+}\longrightarrow \Lambda _{-}
\end{equation*}%
be a map which assigns to each $j\in \Lambda _{+}$ its reflected image $%
rj\in \Lambda _{-}$, i.e. the site symmetric with respect to $P$. The
reflection map $r$ induces a linear morphism $\pi _{P}:$ $\mathfrak{A}%
_{+}\longrightarrow $ $\mathfrak{A}_{-}$, on the abelian algebra $\mathfrak{A%
}_{\pm }$ of bounded function on the configuration space $\Omega _{\pm }=%
\mathbb{R}^{\Lambda _{\pm }}$ given by 
\begin{equation*}
\pi _{P}F\left( \left\{ x_{j}\right\} _{j\in \Lambda _{+}}\right) =F\left(
\left\{ x_{rj}\right\} _{j\in \Lambda _{+}}\right) ~.
\end{equation*}%
By a linear morphism we mean $\pi _{P}\left( FG\right) =\pi _{P}\left(
F\right) ~\pi _{P}\left( G\right) $ is satisfied for any $F,G\in \mathfrak{A}%
_{+}$.

\begin{definition}[Reflection Positivity]
\label{RP}A state $\langle \cdot \rangle _{\nu }$ is said to be a \textit{%
reflection positivity} functional if 
\begin{equation}
\langle F\pi _{P}(F)\rangle _{\nu }\geq 0  \label{pos}
\end{equation}%
holds for all $F$ $\in \mathfrak{A}_{+}$ .
\end{definition}

According to \cite{FILS}, $\langle \cdot \rangle _{\nu _{n}^{(N)}}$ defined
by the expected value with respect to the measure $\nu _{n}^{(N)}$ given by (%
\ref{equ}) with $A=-\Delta \otimes I$, $-\Delta $ the usual Laplacean with
periodic boundary conditions (\ref{laplacean}), is a reflection positivity
functional. Once (\ref{pos}) holds, we have the Schwarz inequality 
\begin{equation}
\langle FG\pi _{P}(FG)\rangle _{\nu }\leq \langle F\pi _{P}(F)\rangle _{\nu
}^{1/2}\langle G\pi _{P}(G)\rangle _{\nu }^{1/2}~.  \label{Sineq}
\end{equation}%
Note that the normalization is unimportant and can be dropped in both sides
of the inequality as long as $n$ and $N$ are kept fixed. Applying (\ref%
{Sineq}) to every plane $P$ which cuts bonds perpendicularly, $\left\langle
\cdot \right\rangle $ satisfies the chessboard inequality (see e.g. \cite%
{FILS})%
\begin{equation}
\left\langle \prod_{j}F_{j}\left( h_{j}\right) \right\rangle _{\nu }\leq
\prod_{j}\left\langle \prod_{i}F_{i}(h_{j})\right\rangle _{\nu }^{1/n}
\label{ineq}
\end{equation}%
where, for each $j$, $F_{j}\left( h,\left\{ x_{i}\right\} _{i\in \Lambda
}\right) =F\left( h,x_{j}\right) $ is a one parameter $h$ family of bounded
function in $\mathbb{R}^{N}$. Note that (\ref{ineq}) has a homogenization
effect.

\noindent \textit{Proof of Theorem \ref{convergence}.} Repeating the steps
of (\ref{In}) and (\ref{I}), we have 
\begin{eqnarray}
I_{n}^{(N)} &=&\frac{1}{S_{N}^{n}}\int_{\mathbb{R}^{n\cdot
N}}\prod_{i=1}^{n}d^{N}x_{i}~\exp \left\{ \frac{-1}{2}\left( \mathbf{x}%
,\left( A-\mu \right) \mathbf{x}\right) \right\}  \notag \\
&=&N^{n/2}\int_{\mathbb{R}_{+}^{n}}\prod_{j=1}^{n}ds_{j}~s_{j}^{N-1}\exp
\left\{ \frac{\mu N}{2}\sum_{j=1}^{n}s_{j}^{2}\right\} ~Z_{n}^{(N)}\left(
1,B\right)  \label{InN}
\end{eqnarray}%
where%
\begin{equation}
B=SJS\otimes I  \label{ZZ}
\end{equation}%
is the coupling matrix $A=J\otimes I$ modified by a matrix $S=\text{diag}%
\left( s_{1},\ldots ,s_{n}\right) $ with the $n$--vector $\mathbf{s}=\left(
s_{1},\ldots ,s_{n}\right) $ in the diagonal. Applying the chessboard
inequality (\ref{ineq}) to this nonhomogeneous partition function (see \cite%
{KZ}), yields%
\begin{equation*}
Z_{n}^{(N)}\left( 1,B\right) \leq \prod_{j=1}^{n}\left( Z_{n}^{(N)}\left(
s_{j}^{2},A\right) \right) ^{1/n}
\end{equation*}%
and we have an upper bound 
\begin{eqnarray}
I_{n}^{(N)} &\leq &N^{n/2}\int_{\mathbb{R}_{+}^{n}}%
\prod_{j=1}^{n}ds_{j}~s_{j}^{N-1}\exp \left\{ \frac{\mu N}{2}%
\sum_{j=1}^{n}s_{j}^{2}\right\} ~\prod_{j=1}^{n}\left( Z_{n}^{(N)}\left(
s_{j}^{2},A\right) \right) ^{1/n}  \notag \\
&=&\left( \sqrt{N}\int_{0}^{\infty }\frac{ds}{s}~\exp \left\{
N~h_{n}^{(N)}(s)\right\} \right) ^{n}  \label{InN1}
\end{eqnarray}%
where, similarly to (\ref{h}),%
\begin{equation*}
h_{n}^{(N)}(s)=\frac{\mu }{2}s^{2}+\ln s+f_{n}^{(N)}(s^{2})~.
\end{equation*}

We are now looking for a lower bound of (\ref{InN}). Using the $O\left(
N\right) $ symmetry together with Jensen inequality, we have%
\begin{eqnarray}
I_{n}^{(N)} &\geq
&N^{n/2}\int_{[c(1-1/N),c]^{n}}\prod_{j=1}^{n}ds_{j}~s_{j}^{N-1}~Z_{n}^{(N)}%
\left( 1,S\left( J-\mu I\right) S\otimes I\right) ~  \notag \\
&=&\left( NR_{N}^{2}\right) ^{n/2}~\left\langle Z_{n}^{(N)}\left( 1,S\left(
J-\mu I\right) S\otimes I\right) \right\rangle  \notag \\
&\geq &\left( NR_{N}^{2}\right) ^{n/2}~Z_{n}^{(N)}\left( 1,\left\langle
S\left( J-\mu I\right) S\right\rangle \otimes I\right) ~  \label{InN2}
\end{eqnarray}%
where $\left\langle \cdot \right\rangle $ denotes the average with respect
to the product measure $\displaystyle\prod_{j=1}^{n}d\mu _{N}(s_{j})$ with $%
d\mu _{N}(s)=ds~s^{N-1}\chi _{\left[ c(1-1/N),c\right] }(s)/R_{N}$ and $%
c=a^{1/N}\sqrt{\beta N^{1/N}}$. The constant $a$ is here chosen arbitrarily
while it has to be tuned properly for the moment generating function. The
normalization 
\begin{equation*}
\sqrt{N}\beta ^{-N/2}R_{N}=\int_{c(1-1/N)}^{c}ds~s^{N-1}=a\left[ 1-\left( 1-%
\frac{1}{N}\right) ^{N}\right] \longrightarrow a\left( 1-\frac{1}{e}\right)
\end{equation*}%
and the two first moments of $\mu _{N}$,%
\begin{equation*}
\left\langle s\right\rangle =\int sd\mu _{N}(s)=\sqrt{\beta }\frac{N^{1+1/2N}%
}{N+1}\frac{1-\left( 1-1/N\right) ^{N+1}}{1-\left( 1-1/N\right) ^{N}}%
\longrightarrow \sqrt{\beta }
\end{equation*}%
and%
\begin{equation*}
\left\langle s^{2}\right\rangle =\int s^{2}d\mu _{N}(s)=\beta \frac{N^{1+1/N}%
}{N+2}\frac{1-\left( 1-1/N\right) ^{N+2}}{1-\left( 1-1/N\right) ^{N}}%
~\longrightarrow \beta
\end{equation*}%
are the only quantities that contribute to the lower bound. We have%
\begin{equation*}
\left( \mathbf{x},\left\langle S\left( J-\mu I\right) S\right\rangle \otimes
I~\mathbf{x}\right) =\left( \mathbf{x},\left( \left\langle s\right\rangle
^{2}A-\left\langle s^{2}\right\rangle \mu \right) ~\mathbf{x}\right)
\end{equation*}%
which, in view of (\ref{Zn}), implies%
\begin{eqnarray*}
Z_{n}^{(N)}\left( 1,\left\langle S\left( J-\mu I\right) S\right\rangle
\otimes I\right) &=&\exp \left( \frac{1}{2}\mu nN\left\langle
s^{2}\right\rangle \right) Z_{n}^{(N)}\left( \left\langle s\right\rangle
^{2},A\right) \\
&=&\exp \left\{ nN\left( \frac{1}{2}\mu \left\langle s^{2}\right\rangle
+f_{n}^{(N)}\left( \left\langle s\right\rangle ^{2}\right) \right) \right\}
~.
\end{eqnarray*}

Similarly to (\ref{I2}), we have%
\begin{equation*}
I_{n}^{(N)}=\frac{2^{n(N/2-1)}\left( \Gamma (N/2)\right) ^{n}}{N^{n(N-1)/2}%
\sqrt{\det \left( J-\mu I\right) }}~.
\end{equation*}%
Taking the limit $\lim\limits_{n,N\rightarrow \infty }\ln I_{n}^{(N)}/nN$,
in any order someone wishes, of both (\ref{InN1}) and (\ref{InN2}) yields%
\begin{equation*}
\frac{\beta }{2}\mu +\frac{1}{2}\ln \beta +f^{\ast }\left( \beta \right)
\leq -\frac{1}{2}-\frac{1}{2}\mathbb{E}\ln \left( J-\mu \right) \leq \frac{%
\mu }{2}\bar{s}^{2}+\ln \bar{s}+f^{\ast }(\bar{s}^{2})
\end{equation*}%
where%
\begin{equation*}
f^{\ast }(\beta )=\lim_{n,N\rightarrow \infty }f_{n}^{(N)}(\beta )
\end{equation*}%
and $\bar{s}$ is the solution to equation (\ref{nu}).

Now we solve (\ref{nu}) with $\bar{s}^{2}=\beta $ for $\mu =\mu (\beta )$
and replace this function in the previous equation to obtain a lower bound%
\begin{equation*}
f^{\ast }(\beta )\geq f(\beta )
\end{equation*}%
with $f$ given by spherical free energy (\ref{f}).

The other side of the same equation gives the following upper bound%
\begin{equation*}
f^{\ast }\left( \beta \right) \leq -\frac{1}{2}\ln \left( e\beta \right)
+\sup_{\mu <0}\frac{1}{2}\left( -\mathbb{E}\ln \left( J-\mu \right) -\beta
\mu \right)
\end{equation*}%
Note that this inequality holds for any $\mu <0$, so it holds for $\mu $
that gives the least upper bound. The supremum is attained at $\mu =\mu
(\beta )$ that solves (\ref{nu}) with $\bar{s}^{2}=\beta $ and we have the
upper bound%
\begin{equation*}
f^{\ast }(\beta )\leq f(\beta )~
\end{equation*}%
concluding the proof of Kac--Thompson theorem for admissible reflection
positive coupling matrices $A$.

\hfill $\Box $

The above proof is now modified to establish convergence of the moments
generating function.

\medskip

\subsection{Proof of Theorem \protect\ref{convergence1}}

The ratio of partition functions 
\begin{equation}
\Theta _{n}^{(N)}(\beta ,z)=\left\langle \exp \left( z(\mathbf{1},\mathbf{x}%
)\right) \right\rangle _{\nu _{n}^{(N)}}=\frac{Z_{n}^{(N)}(\beta ,z,\mathbf{1%
},A)}{Z_{n}^{(N)}(\beta ,A)}  \label{ThetanN}
\end{equation}%
where%
\begin{equation}
Z_{n}^{(N)}(\beta ,z,\mathbf{1},A)=\int \exp \left\{ \frac{-\beta }{2}\left( 
\mathbf{x},A\mathbf{x}\right) +z\left( \mathbf{1},\mathbf{x}\right) \right\}
~\prod_{j=1}^{n}d\sigma _{0}^{(N)}\left( x_{j}\right)   \label{ZnN}
\end{equation}%
with $\mathbf{1}=\left( 1/\sqrt{nN}\right) \left( 1,\ldots ,1\right) $,
generates the moments of $X_{nN}=\left( 1/\sqrt{nN}%
\right) \displaystyle\sum_{i,j}x_{i,j}$. We combine the procedure of Section %
\ref{SMG} with the proof of Theorem \ref{convergence} to evaluate this
ratio. Combining (\ref{Kn}) with (\ref{InN1}) yields%
\begin{eqnarray}
K_{n}^{(N)} &=&\frac{\beta ^{nN/2}}{S_{N}^{n}}\int_{\mathbb{R}^{n\cdot
N}}\prod_{i=1}^{n}d^{N}x_{i}~\exp \left\{ \frac{-1}{2}\left( \mathbf{x}%
,\left( A-\kappa \right) \mathbf{x}\right) +z\left( \mathbf{1},\mathbf{x}%
\right) \right\}   \notag \\
&=&\frac{2^{n(N/2-1)}\left( \Gamma (N/2)\right) ^{n}}{N^{n(N-1)/2}\sqrt{\det
\left( J-\kappa I\right) }}\exp \left\{ \frac{-z^{2}}{2\kappa }\right\}  
\notag \\
&\leq &\left( \sqrt{N}\int_{0}^{\infty }\frac{ds}{s}~\exp \left\{
N~g_{n}^{(N)}(s)\right\} \right) ^{n}  \label{KnN}
\end{eqnarray}%
with%
\begin{equation*}
g_{n}^{(N)}(s)=\frac{\kappa }{2}s^{2}+\ln s+f_{n}^{(N)}(s^{2},s)
\end{equation*}%
where the free energy

\begin{equation*}
f_{n}^{(N)}\left( \beta ,z\right) =\frac{1}{nN}\ln Z_{n}\left( \beta ,z;%
\mathbf{1},A\right) ~
\end{equation*}%
is a smooth function of $(\beta ,z)$ such that $f_{n}^{(N)}\left(
s^{2},0\right) =f_{n}^{(N)}\left( s^{2}\right) $ is given by (\ref{fnN}).
Analogously to (\ref{InN2}), we have%
\begin{eqnarray}
K_{n}^{(N)} &\geq &\left( NR_{N}^{2}\right) ^{n/2}~Z_{n}^{(N)}\left(
1,\left\langle S\left( J-\kappa I\right) S\right\rangle \otimes I\right)  
\notag \\
&=&\left( NR_{N}^{2}\right) ^{n/2}\exp \left\{ nN\left( \frac{1}{2}\kappa
\left\langle s^{2}\right\rangle +f_{n}^{(N)}\left( \left\langle
s\right\rangle ^{2},\left\langle s\right\rangle \right) \right) \right\} 
\label{KnN1}
\end{eqnarray}%
with the constant $a$ in normalization $R_{N}$ chosen so that%
\begin{equation*}
\lim_{N\rightarrow \infty }\sqrt{\frac{N}{\beta }}R_{N}=\sqrt{\frac{2\pi }{%
-\left( g_{n}^{(\infty )}\right) ^{\prime \prime }(\beta )}}\frac{1}{\beta }
\end{equation*}%
and, recalling equation (\ref{Itilde}), the difference between the upper (%
\ref{KnN}) and the lower bound (\ref{KnN1}) when the chemical potential $%
\kappa $ is chosen as a function of $\beta $ is $o(1)$.

One concludes from the last two equations (\ref{KnN}) and (\ref{KnN1}) the
following. If $\bar{s}=\bar{s}(\kappa )$ solves the equation (\ref{sbar}),
then 
\begin{equation*}
\left( \frac{\beta }{2}\kappa +\frac{1}{2}\ln \beta +f_{n}^{(N)}\left(
\beta ,z\right) \right) <-\frac{1}{2}\left( 1+\mathbb{E}\ln \left( J-\kappa
\right) \right) -\frac{z^{2}}{2\kappa }+O(n)<\left( \frac{\kappa }{2}\bar{s%
}^{2}+\ln \bar{s}+f_{n}^{(N)}(\bar{s}^{2},\bar{s})\right) 
\end{equation*}%
holds provided $n$ and $N$ sufficiently large and $\kappa =\kappa
_{n,N}\left( \beta ,z\right) $ solves%
\begin{equation*}
\beta =~\mathbb{E}\frac{1}{-\Delta -\kappa }+\frac{z^{2}}{nN\kappa ^{2}}~
\end{equation*}%
for some $O(n)$ constant independent of $z$. Substituting $\exp \left(
nNf_{n}^{(N)}(\beta ,z)\right) $ in the numerator and denominator (with $z=0$%
) of (\ref{ThetanN}), gives%
\begin{equation*}
\Theta _{n}^{(N)}(\beta ,z)=\exp \left( -nN\frac{\beta }{2}\left( \kappa
-\mu \right) -\frac{nN}{2}\mathbb{E}\ln \frac{J-\kappa }{J-\mu }-\frac{z^{2}%
}{2\kappa }\right) \left( 1+o\left( \frac{1}{N}\right) \right) 
\end{equation*}%
and this implies 
\begin{equation*}
\ln \Theta (\beta ,z)=\lim_{n,N\rightarrow \infty }\ln \Theta
_{n}^{(N)}(\beta ,z)=\frac{-z^{2}}{2\mu (\beta )}
\end{equation*}%
for every $\beta <\beta _{c}$, uniformly in compacts of $z$.

\hfill $\Box $

\begin{remark}
\item 
\begin{enumerate}
\item In order to prove that the theorem holds for the hierarchical
Laplacean we need only to prove that the functional defined by measure (\ref%
{equ}) satisfies reflection positivity.
\end{enumerate}
\end{remark}

\subsection{Reflection Positivity for the Hierarchical Laplacean}

The $O(N)$ Heisenberg hierarchical measure has been shown to satisfy
reflection positivity by Watanabe who considered the model originally
proposed by Dyson with $L^{d}=2$. We extend his proof to the general case $%
L>1$, $d\geq 1$.

Let $\mathbf{\theta }=\left( \theta _{1},\ldots ,\theta _{K}\right) $, $%
\theta _{k}\in \left\{ 0,1\right\} $, be the binary representation of a
point $i$ $\in \left\{ 0,1,\ldots ,2^{K}-1\right\} $. With respect to a
reflection plane $P_{k}$ at the $k$--th hierarchy the map $r$ given by 
\begin{equation*}
\left( r\mathbf{\theta }\right) _{l}=\left\{ 
\begin{array}{lll}
\theta _{l} & \mathrm{if} & l\neq k \\ 
1-\theta _{l} & \mathrm{if} & l=k%
\end{array}%
\right. \ \ 
\end{equation*}%
acts as exchanging each pair of consecutive blocks of size $2^{k-1}$ indexed
by $\mathbf{\tau }=\left( \theta _{k+1},\ldots ,\theta _{K}\right) $:%
\footnote{%
Each pair of consecutive blocks would be reflected : $\left\{ i_{1},\ldots
,i_{2^{k-1}},j_{1},\ldots ,j_{2^{k-1}}\right\} \longrightarrow \left\{
j_{2^{k-1}},\ldots ,j_{1},i_{2^{k-1}},\ldots ,i_{1}\right\} $ if $\left( r%
\mathbf{\theta }\right) _{l}=1-\theta _{l}$ holds for $l\leq k$. We use the
exchange operaction for simplicity.} $\left\{ i_{1},\ldots
,i_{2^{k-1}},j_{1},\ldots ,j_{2^{k-1}}\right\} \longrightarrow \left\{
j_{1},\ldots ,j_{2^{k-1}},i_{1},\ldots ,i_{2^{k-1}}\right\} $.

For the $d$--dimensional lattice, a point $i$ is represented by $\mathbf{%
\theta }=\left( \theta _{1},\ldots ,\theta _{K}\right) $ where $\theta _{k}$
takes values in a box $B_{L}=\left\{ 0,1,\ldots ,L-1\right\} ^{d}$ with
periodic boundary conditions: $\theta _{k}=\left( \theta _{k,1},\ldots
,\theta _{k,d}\right) $ with $\theta _{k,j}$ mod $L\in \left\{ 0,1,\ldots
,L-1\right\} $. As in the previous subsection, for each hierarchy $k$ a
reflection plane $P$ is chosen perpendicular to coordinate axes cutting
bonds, but not sites, of $\left\{ 0,1,\ldots ,L-1\right\} ^{d}$ dividing
this box into two disjoint halves $B_{L}=B_{L}^{+}\cup B_{L}^{-}$. With
respect to a reflection plane $P$ at $k$--th hierarchy the map $r$ assigns
to each $\mathbf{\theta }$ its \textquotedblleft reflected
image\textquotedblright\ $r\mathbf{\theta }$ with $\left( r\mathbf{\theta }%
\right) _{l}=\theta _{l}$ if $l\neq k$ and $\left( r\mathbf{\theta }\right)
_{k}\in B_{L}^{\pm }$ if $\theta _{k}\in B_{L}^{\mp }$.

Let 
\begin{equation*}
\Lambda _{\pm }=\left\{ \mathbf{\theta }:\theta _{k}\in B_{L}^{\pm }~\right\}
\end{equation*}%
be the partition of $\Lambda _{K}=\left\{ 0,\ldots ,L^{K}-1\right\} ^{d}$
into two halves according to a plane $P$ at hierarchy $k$ and let $\mathcal{P%
}_{\pm }$ denote the set of polynomials in $x_{\mathbf{\theta }}$, $\mathbf{%
\theta }\in \Lambda _{\pm }$. The reflection map $r$ induces a linear
morphism $\pi _{P}:$ $\mathcal{P}_{+}\longrightarrow \mathcal{P}_{-}$, given
by 
\begin{equation*}
\pi _{P}F\left( \left\{ x_{\mathbf{\theta }}\right\} _{\mathbf{\theta }\in
\Lambda _{+}}\right) =F\left( \left\{ x_{r\mathbf{\theta }}\right\} _{%
\mathbf{\theta }\in \Lambda _{+}}\right) ~.
\end{equation*}

To prove reflection positivity of the $O(N)$ Heisenberg hierarchical
measure, its enough to show that $-\Delta $ given by (\ref{J}) is a
reflection positivity interaction. For this, let $y=B^{l}x\in \mathbb{R}%
^{\Lambda _{K-l}}$ and notice that $y_{\mathbf{\tau }}\in \mathcal{P}_{\pm }$
according to whether $\theta _{k}\in B_{L}^{\pm }$ are coordinate of $%
\mathbf{\tau }=\left( \theta _{l+1},\ldots ,\theta _{K}\right) $ for $l<k$.
If $l\geq k$, then $y$ can be decomposed as 
\begin{equation*}
y=y^{+}+y^{-}\ ,\ \ y^{\pm }\in \mathcal{P}_{\pm }
\end{equation*}%
with $y^{-}=\pi _{P}y^{+}$. We thus have%
\begin{eqnarray*}
\left( y,y\right) _{\Lambda _{K-l}} &=&\sum_{\mathbf{\tau }:\theta _{k}\in
B_{L}^{+}}\left\vert y_{\mathbf{\tau }}\right\vert ^{2}+\sum_{\mathbf{\tau }%
:\theta _{k}\in B_{L}^{-}}\left\vert y_{\mathbf{\tau }}\right\vert ^{2} \\
&=&\sum_{\mathbf{\tau }:\theta _{k}\in B_{L}^{+}}\left\vert y_{\mathbf{\tau }%
}\right\vert ^{2}+\sum_{\mathbf{\tau }:\theta _{k}\in B_{L}^{+}}\left\vert
\pi _{P}y_{\mathbf{\tau }}\right\vert ^{2} \\
&=&\sum_{\mathbf{\tau }:\theta _{k}\in B_{L}^{+}}\left\vert y_{\mathbf{\tau }%
}\right\vert ^{2}+\pi _{P}\sum_{\mathbf{\tau }:\theta _{k}\in
B_{L}^{+}}\left\vert y_{\mathbf{\tau }}\right\vert ^{2}\equiv \left\Vert
y\right\Vert _{+}^{2}+\pi _{P}\left\Vert y\right\Vert _{+}^{2}
\end{eqnarray*}%
by the morphism property, if $l<k$ and%
\begin{eqnarray*}
\left( y,y\right) _{\Lambda _{K-l}} &=&\sum_{\mathbf{\tau }}\left\vert y_{%
\mathbf{\tau }}^{+}+y_{\mathbf{\tau }}^{-}\right\vert ^{2} \\
&=&\sum_{\mathbf{\tau }}\left\vert y_{\mathbf{\tau }}^{+}\right\vert
^{2}+\sum_{\mathbf{\tau }}\left\vert y_{\mathbf{\tau }}^{-}\right\vert
^{2}+2\sum_{\mathbf{\tau }}y_{\mathbf{\tau }}^{+}y_{\mathbf{\tau }}^{-} \\
&\equiv &\left\Vert y^{+}\right\Vert ^{2}+\pi _{P}\left\Vert
y^{+}\right\Vert ^{2}+2\sum_{\mathbf{\tau }}y_{\mathbf{\tau }}^{+}~\pi
_{P}y_{\mathbf{\tau }}^{+}
\end{eqnarray*}%
and these together with (\ref{J}), according to \cite{FILS}, imply that $%
-\Delta $ is a reflection positivity interaction.


\begin{thebibliography}{FILS}
\bibitem[BK]{BK} T. H. Berlin and M. Kac. \textquotedblleft The Spherical
Model of a Ferromagnet\textquotedblright , Phys. Rev. \textbf{86}, 821-835
(1952)


\bibitem[D]{D} Freeman J. Dyson. \textquotedblleft Existence of a
Phase--Transition in a One--Dimensional Ising Ferromagnet\textquotedblright
, Commun. Math. Phys. \textbf{12}, 91-107 (1969)

\bibitem[Da]{Da} Philip J. Davis. \textquotedblleft Circulant
Matrices\textquotedblright , Chelsea Publishing, N.Y. (1979)




\bibitem[F]{F} Giovanni Felder. \textquotedblleft Renormalization group in
the local potential approximation\textquotedblright , Commun. Math. Phys. 
\textbf{111}, 101-121 (1987)

\bibitem[FILS]{FILS} J\"{u}rg Fr\"{o}hlich, Robert Israel, Elliot H. Lieb
and Barry Simon. \textquotedblleft Phase Transitions and Reflection
Positivity: I. General Theory and Long Range Lattice
Models\textquotedblright , Commun. Math. Phys. \textbf{62}, 1-34 (1978)


\bibitem[GK]{GK} K. Gawedzki and A. Kupiainen. \textquotedblleft
Non--Gaussian Fixed Points of the Block Spin Transformation. Hierarchical
Model Approximation, Commun. Math. Phys. \textbf{89}, 191-220 (1983)


\bibitem[HHW]{HHW} Takashi Hara, Tetsuya Hattori and Hiroshi Watanabe.
\textquotedblleft Triviality of Hierarchical Ising Model in Four
Dimensions\textquotedblright , Commun. Math. Phys. \textbf{220}, 13-40 (2001)

\bibitem[K]{K} Yuri V. Kozitsky. \textquotedblleft Hierarchical
Ferromagnetic Vector Spin Model Possessing the Lee--Yang Property.
Thermodynamic Limit at the Critical Point and Above\textquotedblright ,
Journ. Stat. Phys. \textbf{87}, 799-820 (1997)



\bibitem[Kr]{Kr} Evgenij Kritchevski. \textquotedblleft Spectral
Localization in the Hierarchical Anderson Model\textquotedblright , Proc.
Amer. Math. Soc. \textbf{135}, 1431-1440 (2007)


\bibitem[KKPS]{KKPS} A. M. Khorunzhy, B. A. Khoruzhenko, L. A. Pastur and M.
V. Shcherbina. \textquotedblleft The Large--$n$ Limit in Statistical
Mechanics and the Spectral Theory of Disordered Systems\textquotedblright\
in \textit{Phase Transitions} Vol. 13, Domb and J. Lebowitz ed. (1992)

\bibitem[KT]{KT} Mark Kac and Colin J. Thompson. \textquotedblleft Spherical
Model and Infinite Spin Dimensionality Limit\textquotedblright , Phys.
Norvegica, vol. \textbf{5}, 163-168 (1971)

\bibitem[KZ]{KZ} H. Kunz and G. Zumbach. \textquotedblleft Phase Transition
in a Nematic $N$--Vector Model: The Large $N$ Limit\textquotedblright , J.
Phys. \textbf{A25}, 6155-6162 (1992)

\bibitem[L]{L} Paulo Cupertino Lima. \textquotedblleft Renormalization Group
Fixed Points in the Local Potential Approximation for $d\geq 3$%
\textquotedblright , Commun. Math. Phys. \textbf{170}, 529-539 (1995)


\bibitem[MFH]{MFH} D. H. U. Marchetti, P. A. Faria da Veiga and T. R. Hurd.
\textquotedblleft The $1/N$--Expansion as a Perturbation about the Mean
Field Theory: A One--Dimensional Fermi Model \textquotedblright , Commun.
Math. Phys. \textbf{179}, 623-646 (1996)

\bibitem[MCG]{MCG} D. H. U. Marchetti, William R. P. Conti and Leonardo F.
Guidi. \textquotedblleft Hierarchical Spherical Model from a Geometric Point
of View \textquotedblright , Preprint (2007)

\bibitem[N]{N} Charles M. Newman. \textquotedblleft Inequalities for Ising
Models and Field Theories which Obey the Lee--Yang Theorem\textquotedblright
, Commun. Math. Phys. \textbf{41}, 1-9 (1975)

\bibitem[P]{P} J. Fernando Perez. \textquotedblleft The Role of Gaussian
Domination and Sum Rules in Phase Transitions - An Unpedagogical
Introduction\textquotedblright , Rev. Bras. de F\'{\i}s. \textbf{10},
293-311 (1980)

\bibitem[PP]{PP} Cl\'{a}udio Paiva and J. Fernando Perez. \textquotedblleft
A Hierarchical Model for Random Walks in Random Media\textquotedblright ,
Journ. Stat. Phys. \textbf{71}, 435-452 (1993)

\bibitem[PWH]{PWH} J. F. Perez, W. F. Wreszinski and J. L. van Hemmen.
\textquotedblleft The mean spherical model in a random external field and
the replica method \textquotedblright , Journ. Stat. Phys. \textbf{35},
89--98 (1984)

\bibitem[W]{W} Hiroshi Watanabe. \textquotedblleft Triviality of
Hierarchical $O(N)$ Spin Model in Four Dimensions with Large $N$%
\textquotedblright\ Journ. Stat. Phys. \textbf{115}, 1669-1713 (2004).

\bibitem[Z]{Z} Gil Zumbach. \textquotedblleft The Renormalization Group in
the Local Potential Approximation and its Applications to the $O(n)$
Model\textquotedblright , Nucl. Phys. \textbf{B413}, 754-770 (1994)
\end{thebibliography}
\end{document}